\documentclass[12pt,a4paper,oneside,ngerman]{article}
\pdfoutput=1
\usepackage[T1]{fontenc}
    \usepackage[english,main=ngerman]{babel}
\usepackage[utf8]{inputenc}

\title{Besondere Anforderungen des automatisierten Fahrens an den Entwurf}
\author{{Robert Graubohm} und {Markus Maurer}}
\date{%
{Technische Universität Braunschweig},\\
{Institut für Regelungstechnik}%
}
\usepackage{csquotes}
\usepackage[backend=biber,autolang=hyphen,url=true,maxbibnames=99,maxitems= 99,giveninits=true,maxcitenames=2]{biblatex} 

\DeclareNameAlias{default}{family-given}
   \DeclareFieldFormat*{url}{}
   \DeclareFieldFormat[misc]{url}{\mkbibacro{URL}\addcolon\space\url{#1}}
   \DeclareFieldFormat*{urldate}{}
   \DeclareFieldFormat[misc]{urldate}{\mkbibparens{\bibstring{urlseen}\space#1}}
\PassOptionsToPackage{hyphens}{url}
\AtEveryBibitem{\clearlist{language}\clearfield{note}}
\AtEveryBibitem{\clearfield{isbn}}
\usepackage{graphicx}
\usepackage{hyperref}
\usepackage{hologo}
\usepackage{listings}
\renewcommand*{\mkbibnamefamily}[1]{\leavevmode\begingroup\kern0pt #1\endgroup} 
\begin{document}

\pagestyle{empty}
\noindent
{\Large  Veröffentlicht in Winner, H., Dietmayer, K., Eckstein, L., Jipp, M., Maurer, M., Stiller, C. (Hrsg.) \emph{Handbuch Assistiertes und Automatisiertes Fahren}, 2024}. \\ \\ 
{\Large DOI: \href{https://doi.org/10.1007/978-3-658-38486-9\_45}{10.1007/978-3-658-38486-9\_45}} \\ \\ 
Zu zitieren als:

\vspace{0.1cm}
\noindent\fbox{%
    \parbox{\textwidth}{%
        Graubohm,~R. \& Maurer,~M. (2024). \glqq{}Besondere {Anforderungen} des automatisierten {Fahrens} an den {Entwurf}\grqq{}. In: \emph{Handbuch {Assistiertes} und {Automatisiertes} {Fahren}}. Hrsg. von Winner,~H., Dietmayer,~K., Eckstein,~L., Jipp,~M., Maurer,~M. \& Stiller,~C. 4.~Aufl. ATZ/""MTZ"=Fachbuch. Wiesbaden: Springer, S.~1229--1251. DOI: {10.1007/978-3-658-38486-9\_45}.
    }%
}
\vspace{2cm}

\noindent\begin{minipage}{\textwidth}
\hologo{BibTeX}:
\footnotesize
\begin{lstlisting}[frame=single]
@incollection{winner_besondere_2024,
	address = {Wiesbaden},
	title = {Besondere {Anforderungen} des automatisierten {Fahrens} an den {Entwurf}},
	isbn = {978-3-658-38485-2 978-3-658-38486-9},
	url = {https://link.springer.com/10.1007/978-3-658-38486-9_45},
	language = {de},
	booktitle = {Handbuch {Assistiertes} und {Automatisiertes} {Fahren}},
	publisher = {Springer},
	author = {Graubohm, Robert and Maurer, Markus},
	editor = {Winner, Hermann and Dietmayer, Klaus and Eckstein, Lutz and Jipp, Meike and Maurer, Markus and Stiller, Christoph},
	year = {2024},
	doi = {10.1007/978-3-658-38486-9_45},
	series = {ATZ/MTZ-Fachbuch},
	pages = {1229--1251},
	edition = {4. Aufl},
}
\end{lstlisting}
\end{minipage}

\clearpage

 \maketitle
\thispagestyle{empty}
\begin{abstract}
Die Entwicklung automatisierter Fahrzeuge und Fahrfunktionen stellt eine ausgesprochen komplexe Aufgabe dar, die bereits im Zuge des Systementwurfs die Einbeziehung einer Vielzahl teilweise konfliktärer Interessen und diverser Randbedingungen erfordert. Dieses Kapitel erläutert wichtige Herausforderungen bei Konzeptspezifikationen im Themenfeld des automatisierten Fahrens und stellt ein systematisches Prozessmodell vor, das einen Beitrag zur Erfüllung der besonderen Anforderungen des automatisierten Fahrens an den Entwurf leistet. Darüber hinaus wird die erfolgreiche Durchführung einer strukturierten Konzeptspezifikation für ein automatisiertes Fahrzeugführungssystem beschrieben.
\end{abstract}
\section{Einleitung}
Die (zeitweise) Entbindung der menschlichen Fahrenden von Steuerungs"= und Überwachungsaufgaben stellt eine disruptive Innovation der Fahrzeugentwicklung für den öffentlichen Straßenverkehr dar. Entsprechend groß ist die Herausforderung für die Entwickelnden~-- unter anderem auch bei der Erstellung sowie Anwendung strukturierter und systematischer Prozessmodelle. Im Vergleich mit der Entwicklung klassischer, lediglich mit Assistenzfunktionen ausgestatteter Straßenfahrzeuge werden neue Systeme und Prozesse benötigt, um den Ansprüchen höherer Automatisierungsgrade gerecht zu werden. Insbesondere müssen Lösungen identifiziert werden, die einen Einsatz automatisierter Fahrzeuge und Fahrfunktionen (SAE Level~3 und höher~\cite{sae_international_sae_2021}) in einer offenen Welt ermöglichen. Bisher wird im Zuge der Entwicklung von Assistenzsystemen im Wesentlichen auf den Menschen als Rückfallebene in kritischen und unvorhergesehenen Situationen gesetzt, was die Anwendbarkeit etablierter Entwurfsmethoden für das automatisierte Fahren entscheidend einschränkt.

Die Spezifikation eines belastbaren Satzes von Anforderungen an das System im Sinne einer Funktionsspezifikation kann als zentraler Bestandteil eines Entwurfs aufgefasst werden. Während die abstrakte Beschreibung des angestrebten Funktionsumfangs hierbei unter Umständen trivial erscheint, ist die umfängliche Anforderungsanalyse ein schwer beherrschbares komplexes Problem einer frühen Entwicklungsphase. Insbesondere werden bereits im Zuge initialer Anforderungsdefinitionen Entscheidungen über angestrebte Rückfallstrategien und Sicherheitsmechanismen getroffen, die wesentliche Auswirkungen auf das Systemverhalten in unvorhergesehenen Situationen und an Systemgrenzen haben.

Die zentrale Herausforderung für die Einführung neuer Systeme für das automatisierte Fahren ist die Freigabe des Entwickelten sowohl durch Verantwortliche in herstellenden Unternehmen als auch durch die erforderlichen Prüfstellen. Voraussetzung dafür ist eine überzeugende Sicherheitsargumentation, die den Nachweis der Berücksichtigung relevanter Gesetze und Normen beinhaltet (s.~\cite{winner_testkonzepte_2024}). Ein Nachweis der vollständigen Abwesenheit jedweder Residualrisiken scheint hingegen im Zuge der Freigabe automatisierter Fahrfunktionen unerreichbar~\cite{maurer_hochautomatisiertes_2018}, zumindest solange die Fahrzeuge in dem heute präsenten Verkehrsraum operieren. Umso wichtiger wird im Zusammenhang mit fahrer*innen\-lo\-sen Fahrzeugen ein überzeugendes Sicherheitskonzept, welches Risiken identifiziert, benennt und bestmöglich adressiert. Die Grundlagen des Sicherheitskonzepts und somit das Fundament der Freigabeargumentation werden im Zuge des Systementwurfs geschaffen, der in diesem Kapitel näher erläutert wird. In \autoref{sec2} wird zunächst auf die Definition des Begriffs Entwurf im Kontext mechatronischer Systementwicklung eingegangen. Anschließend werden in \autoref{sec3} besondere Herausforderungen erläutert, die während dieser Entwicklungsphasen im Anwendungsfall des automatisierten Fahrens auftreten. Auf Basis identifizierter Herausforderungen wird in \autoref{sec4} ein Referenzentwicklungsprozess für automatisierte Fahrfunktionen vorgestellt und fachlich eingeordnet. Schließlich folgt eine Einführung in die Anwendung des strukturierten Entwurfsmodells im Zuge des abgeschlossenen Forschungsprojekts aFAS in \autoref{sec5}.

\section{Entwurf mechatronischer Systeme}
\label{sec2}
In der Literatur wird ein \emph{Entwurf} mechatronischer Systeme verschiedenartig aber nicht grundsätzlich widersprüchlich beschrieben. Insofern dienen in Fachbüchern und Normen verwendete Definitionen an dieser Stelle dazu, eine konsistente Definition für dieses Kapitel abzuleiten und auf dieser Basis den Entwurf in einen vollständigen Entwicklungsprozess einzuordnen.

Eine häufig angeführte Beschreibung des Entwurfs als Teil einer Produktentwicklung stammt aus der Theorie des \foreignlanguage{english}{Axiomatic Design}. Im \foreignlanguage{english}{Axiomatic Design} spiegelt der \emph{Entwurf} das Ergebnis eines Zuordnungsproblems von Lösungen auf Anforderungen wider~\cite{suh_axiomatic_2001}. Die Zuordnung erstreckt sich dabei über vier Domänen: Kundendomäne, funktionale Domäne, physische Domäne und Prozessdomäne. Dabei werden aus Kundenwünschen funktionale Anforderungen, Design Parameter und schließlich Prozessvariablen abgeleitet. Funktionale Anforderungen werden dabei als der minimale Satz unabhängiger Anforderungen definiert, der die Entwurfsziele vollständig wiedergibt~\cite[14]{suh_axiomatic_2001}. Außerdem werden Restriktionen zusammen mit den funktionalen Anforderungen erfasst, die den Lösungsraum für weitere Designentscheidungen einschränken können. In diesem Zusammenhang wird die Herausforderung geschildert, Kundenwünsche in \emph{lösungsneutrale} funktionale Anforderungen zu übersetzen, um keine Designentscheidungen vorwegzunehmen und dadurch die Möglichkeit von Innovation zu beeinträchtigen~\cite[15]{suh_axiomatic_2001}.

Die VDI"=Richtlinie~2206~\cite{verein_deutscher_ingenieure_entwicklungsmethodik_2004} übernimmt die Definition des Terms \emph{Entwurf} im Kontext der mechatronischen Systementwicklung aus Beschreibungen des Syntheseschritts durch \citeauthor{daenzer_systems_1999}:
  \begin{quote}
    \glqq{}die \glq{}Erahnung\grq{} eines Ganzen, eines Lösungskonzepts, das Erkennen bzw. \glq{}Finden\grq{} der dazu erforderlichen Lösungselemente und das gedankliche, modellhafte Zusammenfügen und Verbinden dieser Elemente zu einem tauglichen Ganzen\grqq{}~\cite[158]{daenzer_systems_1999}.
  \end{quote}
  
  Insofern stellt der Entwurf grundsätzlich einen frühen Teilschritt der gesamten Entwicklung eines Produktes dar, der entgegen anderer Definitionen aus dem Maschinenbau auch das Konzipieren umfasst und in dem technischen System zu einer Konkretisierung von Anforderungen führt~\cite[Abschn.~3.1.2]{verein_deutscher_ingenieure_entwicklungsmethodik_2004}. Zum Resultat eines erfolgten Entwurfs führt die Richtlinie aus:
  \begin{quote}
    \glqq{}Ergebnis des Systementwurfs ist das abgesicherte Konzept eines mechatronischen Systems. Darunter wird die prinzipiell festgelegte und durch \emph{Verifikation} und \emph{Validierung} überprüfte Lösung verstanden.\grqq{}~\cite[8]{verein_deutscher_ingenieure_entwicklungsmethodik_2004}
  \end{quote}
  
  Den vorgestellten Definitionen folgend soll im Weiteren durch den Entwurf grundsätzlich eine Konkretisierung eines abstrakten Entwicklungsziels (üblicherweise aufgrund eines identifizierten Kund*innen\-nut\-zens) durch konkrete technische Lösungsansätze verstanden werden. Die hervorgebrachte Lösung soll auf einem umfassenden Satz funktionaler sowie nicht"=funktionaler Anforderungen basieren und bereits durch erste Studien, Prototypen oder Muster validiert sein. In den folgenden Abschnitten wird außerdem erläutert, dass ein erfolgreicher Systementwurf im Kontext des automatisierten Fahrens darüber hinaus von der Spezifikation eines zwar vorläufigen aber belastbaren Sicherheitskonzepts abhängt. Die finale Ausgestaltung und Herstellungsplanung eines Serienprodukts sowie dessen Absicherung und Freigabe für Herstellung und Vertrieb sind dem Entwurf nachgelagerte Schritte einer Produktentwicklung.

Dabei ist eine Einordung des Entwurfs in die Phasen des Sicherheitslebenszyklus nach ISO~26262 (s.~\cite{winner_sicherheit_2024-1}) herausfordernd: Die Zielsetzung eines belastbaren funktionalen Konzepts teilt sich der Entwurf mit der Konzeptphase nach ISO~26262, die in Teil~3 der Norm~\cite{international_organization_for_standardization_iso_2018} adressiert wird. Entsprechend gilt es, im Zuge des Entwurfs elektrischer und elektronischer Fahrzeugsysteme den Forderungen an Prozess und Arbeitsprodukte der Konzeptphase besondere Beachtung zu schenken. Sobald allerdings zur Konzeptvalidierung Lösungen bereits prototypisch umgesetzt und erprobt werden, werden innerhalb des Entwurfs auch spätere Phasen des Sicherheitslebenszyklus nach ISO~26262 (Entwicklung auf Systemebene, Hardwareebene und Softwareebene) in frühen Stadien durchlaufen.

\section{Herausforderungen bei Konzeptspezifikationen im Themenfeld des automatisierten Fahrens}
\label{sec3}
Die Entwicklung umfangreicher mechatronischer Systeme~-- wie beispielsweise automatisierter Fahrfunktionen und Fahrzeuge~-- weist substanzielle Herausforderungen auf, die insbesondere aus der Komplexität der technischen Systeme und der Vielfalt der Situationen in ihren Einsatzdomänen resultieren. Der frühzeitigen Berücksichtigung und Adressierung dieser Herausforderungen bereits im Entwurfsstatus wird dabei besondere Bedeutung für eine erfolgreiche Konzeptspezifikation beigemessen. So führt die VDI"=Richtlinie~2206 aus, dass die Problematik von Wechselwirkungen einer Vielzahl von Komponenten, die das Verhalten des Gesamtsystems beeinflussen, \glqq{}bereits in der frühen Phase des Entwurfs zu berücksichtigen [ist]\grqq{}~\cite[4]{verein_deutscher_ingenieure_entwicklungsmethodik_2004}.

Zentral ist hierbei, die Thematik emergenter Eigenschaften zu beachten, die ein entworfenes Gesamtsystem ausprägt, ohne dass sie direkt auf Eigenschaften einzelner Elemente des Gesamtsystems zurückzuführen sind~\cite[97]{rausand_risk_2020}. Das Konzept emergenter Eigenschaften eines Systems, die beabsichtigt oder unbeabsichtigt aus einer Interaktion von Komponenten resultieren, wird als wesentlicher Aspekt der Systementwicklung gesehen~\cite[5]{dick_requirements_2017}. Dabei ist die Sicherheit von Systemen eine zentrale emergente Eigenschaft~\cite[64]{leveson_engineering_2012}, die zugleich von höchster Relevanz im Kontext des automatisierten Fahrens ist. Von einer Konzeptspezifikation wird entsprechend erwartet, dass sie eine Basis dafür schafft, dass der Entwicklungsgegenstand emergente Eigenschaften bzw. emergentes Systemverhalten in einer angestrebten Weise ausbildet.

Ein weiterer Aspekt der Komplexität mechatronischer Entwicklungsprojekte ist der interdisziplinäre Charakter des Produkts und seiner Konzipierung. Das Ziel des mechatronischen Systementwurfs wird zum Beispiel durch die VDI"=Richtlinie~2206 als \glqq{}Festlegung eines domänenübergreifenden Lösungskonzepts\grqq{} erläutert, was die Notwendigkeit einer interdisziplinären Zusammenarbeit und Kommunikation herausstellt~\cite[Abschn.~3.1.2]{verein_deutscher_ingenieure_entwicklungsmethodik_2004}. Dabei wird angeführt, dass eine systematische Abstimmung der Domänen in frühen Entwicklungsphasen kosten"= und zeitintensive Iterationen vermeidet, die durch sequenzielles Vorgehen ohne Abstimmung entstehen~\cite[Abschn.~2.4]{verein_deutscher_ingenieure_entwicklungsmethodik_2004}. Der in der Richtlinie referenzierte domänenübergreifende Charakter einer Entwicklung mechatronischer Systeme bezieht sich in Referenz zu Definitionen aus der Literatur auf das Zusammenwirken der Disziplinen Informationstechnik, Maschinenbau und Elektrotechnik~\cite[Abschn.~2.1]{verein_deutscher_ingenieure_entwicklungsmethodik_2004}. Für den Entwurf automatisierter Fahrfunktionen und Fahrzeuge sind die einwirkenden Domänen tatsächlich noch weiterzufassen: Ergonomie (vgl.~\cite{winner_nutzergerechte_2024,winner_human_2024}) und Rechtswissenschaften (vgl.~\cite{winner_allgemeine_2024,winner_rahmenbedingungen_2024}) sind Beispiele für Disziplinen, die zusätzlich frühzeitig in der Konzeptspezifikation berücksichtigt werden müssen.

Die kurzen Innovationszyklen bei Schlüsseltechnologien der Automatisierung der Fahraufgabe sind zugleich Chance und Herausforderung innerhalb des Entwurfs zukünftiger Fahrzeugsysteme. Fortschreitende Entwicklungen und Verbesserungen eingesetzter Sensoren, Steuergeräte und Algorithmen schaffen eine Basis für neuartige anspruchsvolle Funktionsumfänge. Allerdings manifestiert diese Evolution eingesetzter Technologie gleichzeitig die Notwendigkeit, im Zuge von Konzeptspezifikationen des automatisierten Fahrens Vorhersagen über Serienreife sowie Verfügbarkeiten und Kostenniveaus zu machen, deren Validität im Zuge eines fortschreitenden Entwicklungsprozesses kontinuierlich geprüft werden muss. Der populäre Einsatz innovativer Softwareentwicklungsansätze~-- wie des maschinellen Lernens~-- kann darüber hinaus erhebliche Auswirkungen auf die Strukturierung folgender Phasen der Gesamtsystementwicklung haben, die zum Zeitpunkt der Konzeptspezifikation bereits einzuplanen sind. Funktionen, die (teilweise) auf maschinellem Lernen basieren, erfordern einerseits die Verfügbarkeit oder Erzeugung erschöpfender Trainingsdaten und andererseits zusätzliche Testumfänge, um Robustheit und Sicherheit der im Regelfall nicht durch Menschen interpretierbaren Softwarebestandteile zu prüfen.

Die Aspekte der hohen System"= und Entwicklungskomplexität, des interdisziplinären Charakters einer Konzeptspezifikation sowie die fortschreitende Evolution einsetzbarer Technologien führen zu einer besonderen Bedeutung der Anforderungsdefinition für den Entwurf automatisierter Fahrfunktionen und Fahrzeuge. Allgemein wird im Zuge des Entwurfs eines technischen Systems häufig von einer Konkretisierung initial bereits vorliegender abstrakter Anforderungen ausgegangen. So beschreibt die VDI"=Richtlinie~2206, dass ein Entwicklungsauftrag, in dem die Aufgabenstellung präzisiert und in Form von Anforderungen beschrieben wurde, den Ausgangspunkt des Systementwurfs bilde~\cite[Abschn.~3.1.2]{verein_deutscher_ingenieure_entwicklungsmethodik_2004}. Im Bereich des automatisierten Fahrens liegen allerdings vielfältige und häufig konträre Stakeholderinteressen vor, die im Zuge der Konzeptspezifikation harmonisiert werden müssen. Fahrer*innen\-lo\-se Fahrzeuge sollen die Erwartungen Nutzender erfüllen und dabei sehr hohe Sicherheitsniveaus realisieren, sie müssen aber letztlich auch von anderen am Verkehr Teilnehmenden und der Gesellschaft toleriert und akzeptiert werden (vgl.~\cite[159]{ross_safety_2021}). Entsprechend werden auch abstrakte Systemanforderungen im Verlauf eines Entwurfs hinterfragt und müssen oftmals revidiert werden.

Die besondere Einordnung der Konzeptphase und Konzeptspezifikation im Bereich der Automobilindustrie wird auch von der VDI"=Richtlinie~2221 wiedergegeben: In Beispielen verschiedener Entwicklungsprozesse wird hier in der Automobildomäne einschließlich der Zulieferindustrie die Klärung und Präzisierung der Aufgabe als Teil einer umfangreichen Phase der Konzeptentwicklung beschrieben~\cite[Blatt~2]{verein_deutscher_ingenieure_entwicklung_2019}. Im Themenfeld des automatisierten Fahrens stellt die fordernde Definition, Konkretisierung und Revision von Anforderungen tatsächlich einen Hauptaspekt der Konzeptphase dar. Ein Beispiel für diesen Umstand sind Sicherheitsanforderungen, deren Rolle im Entwurfsprozess in \autoref{ssec32} näher erläutert wird. Unabhängig des Abstraktionsgrads kann ein Sicherheitskonzept nur auf Basis vorher definierter Funktionsumfänge erzeugt werden. Gleichzeitig hat es aber erheblichen Einfluss auf die Anforderungsliste des Gesamtsystems: Unter Sicherheitsaspekten können neue Anforderungen erzeugt oder Anpassungen existenter Anforderungen provoziert werden.

Im Zuge der Vorstellung des als Prozessmodell für die Systementwicklung eingeführten V"=Modells~XT wird angeführt, dass durch Sicherheitsanalysen \glqq{}neue funktionale oder nichtfunktionale Anforderungen entstehen können\grqq{}~\cite[192]{hoppner_ag--schnittstelle_2008}, die die Anforderungsfestlegung in einem Lastenheft ergänzen. Durch die Ausführungen wird zwar der dynamische Charakter einer Anforderungsliste in umfangreichen Systementwicklungsprojekten zumindest durch Sicherheitsbetrachtungen aufgezeigt, jedoch wird nicht der rekursive und iterative Charakter der eigentlichen Anforderungsdefinition illustriert, der bei der Spezifikation automatisierter Fahrfunktionen vorliegt. Denn wie zuvor erläutert nehmen Sicherheitsanalysen in der Konzeptphase vielmehr direkten Einfluss auf die eigentliche Spezifikation einer fahrer*innen\-lo\-sen Anwendung und können Anpassungen jeder zuvor dokumentierten Funktionsanforderung erfordern. Auch die von \textcite{dick_requirements_2017} eingeführten Abstraktionsgrade für die Anforderungsanalyse illustrieren zwar die kontinuierliche Erweiterung und Konkretisierung von Anforderungen durch den gesamten Entwurfsprozess, scheinen aber ebenfalls eher von einer sequenziellen Anforderungsdefinition auszugehen. In der Norm ISO~9241"~210 zur menschzentrierten Gestaltung interaktiver Systeme~\cite{deutsches_institut_fur_normung_din_2020} wird hingegen abgebildet, dass der Nutzungskontext eines Systems auf Basis erzeugter Lösungskonzepte iterativ neu analysiert werden muss, was zu neuen Anforderungen und gegebenenfalls wiederum neuen Lösungskonzepten führt. Die Empfehlungen der Norm für Gestaltungsaktivitäten scheinen somit besonders für sicherheitskritische Anwendungen gut übertragbar. In diesem Kontext führen \textcite{sottilare_towards_2020} an, dass die Evaluation des Nutzungskontexts über den gesamten Lebenszyklus eines Produkts fortgesetzt werden sollte, um die Notwendigkeit von Anpassungen auch zu späteren Zeitpunkten identifizieren zu können.

\subsection{Unsicherheiten während der Anforderungsdefinition}
\label{ssec31}
Wie zuvor erläutert erfolgt aus unterschiedlichen Gründen die Anforderungsdefinition während des Entwurfs automatisierter Fahrfunktionen und Fahrzeuge iterativ im Zuge einer ausgeprägten Konzeptphase. Neben der zentralen Herausforderung, Anforderungsanalysen systematisch durchzuführen und dabei erschöpfende und nachvollziehbare Anforderungslisten zu erstellen, spielen im Themenfeld des automatisierten Fahrens auch Unsicherheiten eine große Rolle. Im vorangegangenen Abschnitt wurde der notwendige Umgang mit technologischen Unsicherheiten über zukünftige erreichbare Performanz und Kosten bereits erwähnt. Daneben stellen automatisierte Fahrfunktionen und Fahrzeuge Beispiele disruptiver innovativer Produkte dar, für die Zahlungsbereitschaften und Marktakzeptanz großen Unsicherheiten unterliegen.

Eine wichtige Unsicherheit im Zuge einer Konzeptphase ist, ob die Definition bestimmter Anforderungen tatsächlich zur wunschgemäßen Ausprägung emergenten Verhaltens komplexer Systeme führt. Die Dokumentation des Systementwurfs in verschiedenen Architektursichten ist ein Ansatz zur Beherrschung vorliegender Komplexität (s.~\cite{winner_architektursichten_2024}). Es besteht dabei allerdings die Möglichkeit, dass die Gestaltung einer vorläufigen Systemarchitektur gemäß vorliegender Anforderungen geschieht, das tatsächlich erzeugte Produkt aber von der konzipierten Lösung abweicht (vgl. Ausführungen zum Architekturdrift und zur Architekturerosion in~\cite{winner_architektursichten_2024}).

Im Anwendungsfall der Automatisierung der Fahraufgabe für den öffentlichen Straßenverkehr spielt darüber hinaus insbesondere auch der Umgang mit Unsicherheiten aufgrund möglicher Fehler des entwickelten technischen Systems eine Rolle. Fehler können allgemein dazu beitragen, dass von dem im Zuge der Konzeptspezifikation angestrebten Fahrzeugverhalten im Betrieb deutlich abgewichen wird. Im Kontext des Straßenverkehrs führen solche Abweichungen häufig zu Regelverletzungen und können erhebliche Gefährdungen verursachen. Grundsätzlich ist dabei der Umgang mit systematischen Fehlern eines technischen Systems und der Umgang mit zufälligen Hardwarefehlern zu unterscheiden: Während zufälligen Hardwarefehlern ein zufallsbedingter Ausfall mindestens einer Hardwarekomponente zugrunde liegt, treten systematische Fehler insbesondere aufgrund menschlicher Fehler oder Fehleinschätzungen im Laufe einer Entwicklung und durch Defizite während der Produktion auf (vgl.~\cite[Teil~1, 3.119 \&~3.165]{international_organization_for_standardization_iso_2018}, \cite[30]{rausand_risk_2020}).

Beide Arten von Fehlern haben direkten Bezug zur Behandlung von Unsicherheiten während der Anforderungsdefinition. Systematische Fehler entstehen zum Teil bereits während der Konzeptspezifikation. So können zum Beispiel aus unvollständigen oder unklaren Anforderungen Fehler im entwickelten System resultieren. Die Möglichkeit unvollständiger Anforderungen ist dabei teilweise auch auf weitere während des Entwurfs vorhandene Unsicherheiten zurückzuführen. Zum Beispiel herrscht Unsicherheit über mögliche Ausprägungen und Auftretenswahrscheinlichkeiten einzelner Betriebsszenarien in der vorgesehenen Einsatzumgebung. Selbst in stark eingeschränkten Anwendungsrahmen ergibt sich in der realen Welt grundsätzlich eine beliebig große Anzahl unterscheidbarer Situationen, in denen sich ein automatisiertes Fahrzeug befinden kann und die es prinzipiell beherrschen muss. Dabei existiert zusätzliche Unsicherheit, ob sich andere Verkehrsteilnehmer in Interaktion mit fahrer*innen\-lo\-sen Fahrzeugen anders verhalten, als sie es heute im Straßenverkehr tun.

Sofern die Unvorhersehbarkeit bestimmter Szenarien im Zusammenhang mit den Leistungsgrenzen oder vorhersehbarem Fehlgebrauch einzelner Funktionen eines Fahrzeugs zu unsicherem Verhalten führt, wird gemäß der Norm ISO/""DIS~21448~\cite{international_organization_for_standardization_isodis_2021} von Residualrisiken gesprochen. Dabei gilt der Grundsatz, dass Residualrisiken evaluiert werden müssen und nur bis zu einem bestimmten Umfang akzeptabel sind (vgl.~\cite[Abschn.~6.3]{winner_sicherheit_2024-1}). Residualrisiken können und sollten zwar im Zuge einer Entwicklung mit besonderen Bemühungen reduziert werden, sie können jedoch niemals vollkommen vermieden werden, da mit dem Einsatz automatisierter Fahrzeuge und Fahrfunktionen im öffentlichen Verkehrsraum inhärente Risiken einhergehen~\cite{nolte_supporting_2020}. Die Anforderungsdefinition kann inhärente Risikoumfänge zum Teil durch Anpassung von Funktionsumfängen und vorgesehenen Einsatzgebieten beeinflussen, verbleibende Risiken und ihre ursächlichen Unsicherheiten sollten jedoch den Entwurf begleitend offengelegt werden, um eine Basis für eine öffentliche Debatte über akzeptable Risikoumfänge zu schaffen~\cite{maurer_wirtschaft_2005,maurer_hochautomatisiertes_2018}.

Während systematische Fehler eines entwickelten technischen Systems also nicht allein ursächlich für vorliegende Risiken sind, führen sie doch zu einer unmittelbaren Erhöhung der Residualrisiken, wenn aus ihnen in bestimmten Situationen unsicheres Verhalten resultiert. Entsprechend sollte ein Prozess stets so gestaltet sein, dass durch systematisches Vorgehen und ausreichende Prüfung der Arbeitsergebnisse systematische Fehler vermieden oder durch Tests identifiziert und beseitigt werden. Ein verbreiteter Ansatz zur Vermeidung systematischer Fehler im Zusammenhang mit dem Systementwurf ist die modellbasierte Systementwicklung (\foreignlanguage{english}{Model-based systems engineering}). Ein zentraler Aspekt des Ansatzes ist das Sicherstellen, dass Anforderungen durch umzusetzende Funktionen adressiert werden, indem sie in Modellen miteinander verknüpft werden (vgl.~\cite{wymore_model-based_1993}).

Neben der Vermeidung systematischer Fehler sollten zufälligen Fehlern im Zuge der Anforderungsdefinition klare Grenzwerte zugeordnet werden, da sich sonst prinzipiell beliebig hohe Ausfallraten des Gesamtsystems ergeben können. Wenn in einer Gefährdungsanalyse festgestellt wird, dass bestimmte mögliche Fehler zu bemerkenswerten Risiken führen, werden Grenzwerte durch sogenannte \foreignlanguage{english}{Automotive}"=Sicherheitsintegritätsstufen (ASIL) gemäß ISO~26262 ausgedrückt. Treten bestimmte Ausfallarten häufiger auf als durch das Integritätsniveau erlaubt, birgt der Einsatz des entwickelten Systems in der vorgesehenen Einsatzumgebung unzumutbare Risiken (vgl.~\cite[Teil~5, 9]{international_organization_for_standardization_iso_2018}). ASIL"=Klassifikationen dienen innerhalb der ISO 26262 außerdem dazu, einen notwendigen Umfang von Prozessen und Entwicklungsmethoden zur Vermeidung systematischer Fehler vorzuschreiben (s.~\cite[Abschn.~6.2.2.1]{winner_sicherheit_2024-1}).

\subsection{Konzeptphase als Grundlage der Produktsicherheitsargumentation}
\label{ssec32}
In den vorangegangenen Abschnitten wurde bereits deutlich, dass Sicherheitsaspekte schon in initialen Phasen des Entwurfs automatisierter Fahrfunktionen und Fahrzeuge eine wichtige Rolle spielen. Da die Festlegung notwendiger Sicherheitsstrategien und Sicherheitsmechanismen wesentliche Implikationen für das Systemdesign haben kann, muss sie frühzeitig auf Basis einer Evaluierung möglicher Gefährdungen erfolgen. Diese Gefährdungen resultieren im Wesentlichen aus dem angestrebten Funktionsumfang und den Grenzen der Einsatzumgebung. Ob die Entwicklung bestimmter Systeme weiterverfolgt wird, hängt also auch davon ab, ob ein systemweites Sicherheitskonzept, das mögliche Gefährdungen hinreichend adressiert, umsetzbar scheint.

Deswegen ist das Ziel eines fahrer*innen\-lo\-sen Betriebs von Straßenfahrzeugen auch mit einer Entwicklung mechatronischer Komponenten (Sensoren, Aktuatoren und Steuergeräte) und besonderer Algorithmen verbunden, die nicht unmittelbar der Erfüllung der angestrebten Funktion dienen. Viel eher dienen diese Elemente der Argumentation ausreichender Sicherheit in unbekannten Szenarien oder unter Degradationsbedingungen. Die zuvor erwähnte Umsetzbarkeit von Sicherheitsstrategien ist allerdings neben technischen auch von ökonomischen Aspekten abhängig. So können voraussichtliche Kosten der für die Realisierung einer angestrebten Funktionalität auf einem akzeptablen Sicherheitsniveau notwendigen Hardwarekomponenten durchaus zu einer bedeutenden Anpassung des Entwicklungsziels führen.

Im Zuge des Systementwurfs werden also nicht nur konkrete technische Lösungsansätze für Funktionen, sondern auch Lösungsansätze für Sicherheitsstrategien identifiziert. Die Wirksamkeit und Realisierbarkeit einzelner Strategien kann dabei bereits mit Studien, Prototypen oder Mustern validiert werden. Der Entwurf automatisierter Fahrfunktionen und Fahrzeuge beinhaltet darüber hinaus auch Überlegungen, wie die Sicherheit als emergente Eigenschaft des konzipierten Systems schließlich evaluiert und argumentiert werden soll.

Insofern bildet die frühe Dokumentation eines Sicherheitskonzepts die Basis für eine umfassende und durch Testergebnisse gestützte Sicherheitsargumentation, die für eine Serienfreigabe der Funktion oder des Fahrzeugs erforderlich ist. Eine solche Serienfreigabe erfolgt in jedem Fall durch Ent\-schei\-dungs\-trä\-ger*innen im herstellenden Unternehmen, kann zum Beispiel innerhalb der Europäischen Union aber auch durch externe Prüfgesellschaften vor Markteinführung erforderlich sein. Wesentliche Beiträge des Systementwurfs zur Sicherheitsargumentation sind die Dokumentation
\begin{itemize}
\item rechtlicher und normativer Randbedingungen,
\item angewandter Verfahren zur Gefährdungsidentifikation und deren Ergebnisse,
\item identifizierter Risikoursachen und vorliegender Unsicherheiten über unbekannte Szenarien sowie angestrebter oberer Grenzen für Residualrisiken,
\item konzipierter Strategien zur Vermeidung unzumutbarer Risiken (ausgedrückt in Sicherheitsanforderungen),
\item vorliegender Evidenz der Wirksamkeit eingeführter Sicherheitsstrategien,
\item einer umfassenden Systembeschreibung, die eine Definition der Einsatzdomäne, vorgesehene Sicherheitsmechanismen und eine vorläufige Systemarchitektur beinhaltet,
\item bereits im Zuge der Konzeptspezifikation bekannter qualitativer und quantitativer Residualrisiken sowie
\item notwendiger Verfahren und Methoden zur Sicherstellung der Anforderungskonformität im Zuge der Umsetzung in einer Serienentwicklung sowie durch Tests.
\end{itemize}

Zusammenfassend wurde im \autoref{sec3} erörtert, dass Funktionskonzepte und Funktionsumfänge automatisierter Fahrfunktionen so gestaltet werden müssen, dass mehrere kritische Komponenten bedient werden. Zentrale Beispiele sind
\begin{itemize}
\item ein belastbares Sicherheitskonzept,
\item der Kund*innen\-nut\-zen (insbesondere auch der subjektive Nutzen),
\item die Marktfähigkeit (Kosten/""Nutzen und Kommunikationsaspekte) und
\item die Integration in einem Fahrzeugkonzept (insbesondere technologische Aspekte: Rechenleistung, Energiebedarf, Bauraumbedarf, …).
\end{itemize}

Fahrzeuge, die sich automatisiert im öffentlichen Raum bewegen, weisen im Vergleich zu vielen anderen Systemen sehr diverse Stakeholdergruppen auf: Neben Erwartungen der Mitfahrenden sind insbesondere auch Erwartungen anderer am Verkehr Teilnehmender~-- auch der Nutzer*innen von Fuß"= und Radwegen~-- im Zuge des Entwurfs systematisch zu berücksichtigen. Belastbare Konzeptspezifikationen müssen entsprechend sowohl bei Nutzenden als auch bei anderen am Verkehr Teilnehmenden Akzeptanz finden und zugleich ein gesellschaftlich tolerierbares Sicherheitsniveau aufweisen.

\section{Referenzentwicklungsprozess für automatisierte Fahrzeuge}
\label{sec4}
Ein Referenzentwicklungsprozess leistet einen wichtigen Beitrag zur Überwindung der im \autoref{sec3} erläuterten Herausforderungen bei Konzeptspezifikationen innerhalb eines Entwurfs automatisierter Fahrfunktionen und Fahrzeuge. Die Orientierung an einem strukturierten Entwicklungsprozess hilft, die frühzeitige und systematische Berücksichtigung der zuvor angeführten entscheidenden Aspekte sicherzustellen. Die in diesem Kapitel vorgestellte Prozessstruktur soll also Referenz für die Gestaltung und Überwachung laufender Entwicklungsprojekte sein, die sich in frühen Entwicklungsphasen befinden, also im industriellen Kontext üblicherweise durch Forschung und Vorserienentwicklung behandelt werden.

\subsection{V"=Modell als domänenübergreifendes Prozessmodell}
\label{ssec41}
Eine etablierte Referenz für das Vorgehen bei der Entwicklung komplexer Systeme ist das sogenannte V"=Modell. Das Prozessmodell beschreibt im Kern die Kombination einer \foreignlanguage{english}{Top-Down}"=Dekomposition mit einer \foreignlanguage{english}{Bottom-Up}"=Integration, die in Gestalt des Buchstaben V illustriert wird. Die Ursprünge des Modells liegen in Softwareentwicklungsprojekten der US"=Bundesbehörde für Raumfahrt und Flugwissenschaft NASA~\cite{forsberg_relationship_1991}.

In der VDI"=Richtlinie~2206 wird das V"=Modell als Makrozyklus der mechatronischen Systementwicklung eingeführt, der die Phasen Systementwurf, domänenspezifischer Entwurf und Systemintegration aufweist~\cite[Abschn.~3.1.2]{verein_deutscher_ingenieure_entwicklungsmethodik_2004}.
\begin{quote}
\glqq{}Das V"=Modell beschreibt das generische Vorgehen beim Entwurf mechatronischer Systeme, das fallweise auszuprägen ist.\grqq{}~\cite[29]{verein_deutscher_ingenieure_entwicklungsmethodik_2004}
\end{quote}

\begin{figure}
        \centering
        \includegraphics[width=.789\textwidth]{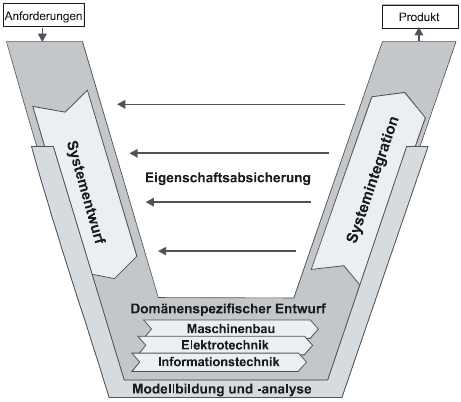}
        \caption{V"=Modell als Makrozyklus~\cite[Abbildung~3-2]{verein_deutscher_ingenieure_entwicklungsmethodik_2004}}
        \label{fig1}
        \end{figure}

Darüber hinaus werden die Phasen des Makrozyklus durch den Prozessbaustein Modellbildung und "~ana\-ly\-se im Sinne einer modellbasierten Systementwicklung flankiert (s. \autoref{fig1}).

Neben dem Potential, die Komplexität mechatronischer Systeme durch ein konsequentes \foreignlanguage{english}{Top-Down}"=Vorgehen bei der Festlegung der Grobstruktur zu beherrschen, betont die Richtlinie auch die Notwendigkeit, den Makrozyklus in Gestalt des V"=Modells iterativ zu durchlaufen.
\begin{quotation}\noindent
\glqq{}Ein komplexes mechatronisches Erzeugnis entsteht in der Regel nicht innerhalb eines Makrozyklus. Vielmehr sind mehrere Durchläufe erforderlich.

In einem ersten Zyklus werden beispielsweise das Produkt funktional spezifiziert, erste \emph{Wirkprinzipien} und/""oder \emph{Lösungselemente} ausgewählt und grobdimensioniert, im Systemkontext auf Konsistenz geprüft und exemplarisch realisiert.\grqq{}~\cite[30]{verein_deutscher_ingenieure_entwicklungsmethodik_2004}
\end{quotation}

Die Möglichkeit mehrerer Durchläufe wird innerhalb der VDI"=Richtlinie~2206 durch verschachtelte V"=Modelle illustriert, worin mit jedem Durchlauf die Produktreife zunimmt (vgl.~\cite[31]{verein_deutscher_ingenieure_entwicklungsmethodik_2004}). Der Makrozyklus teilt sich jedoch die stark vereinfachte Annahme mit vielen anderen Versionen des V"=Modells, dass die Systemanforderungen als Eingangsgrößen für die Systementwicklung zu Beginn vorliegen (s. \autoref{fig1}). Insofern wird in diesem Prozessmodell kein besonderer Fokus auf die Wiedergabe einer systematischen Anforderungsdefinition auf verschiedenen Abstraktionsebenen gelegt. Grafisch erfolgt zudem keine Betonung der sofortigen Rückführung zur Anpassung der Funktionsspezifikation, sobald wesentliche Entwurfskonflikte auftreten, die als unlösbar eingestuft werden können (vgl. \autoref{ssec43}).

\textcite[Abschn.~1.5]{dick_requirements_2017} definieren verschiedene Abstraktionsgrade für Anforderungen und nachvollziehbare Verknüpfungen zwischen den Ebenen innerhalb der \foreignlanguage{english}{Top-Down-Bottom-Up}"=Methodik des V"=Modells: Zunächst werden Anforderungen durch den Entwurfsprozess konkretisiert und erweitert. Im Zuge der Verifikation wird dann auf verschiedenen Abstraktionsniveaus gegen die Anforderungen getestet. Die Autor*innen argumentieren, dass eine Strukturierung des Prozesses die Basis einer nachvollziehbaren Verknüpfung von Anforderungen bildet, die die Voraussetzung für Wirkungsanalysen im Zuge des Änderungsmanagements ist. Wie bereits in \autoref{sec3} beschrieben wird allerdings durch \citeauthor{dick_requirements_2017} eher ein sequenzielles Vorgehen für die Anforderungsspezifikation dargelegt und auch hier kein dynamisch iterativer Prozess abgebildet.

Eine konkrete Weiterentwicklung des Makrozyklus \emph{V"=Modell} unter den besonderen Anforderungen der Systemtechnik wird von \textcite{rios_evolution_2017} vorgestellt. Darin wird insbesondere die Integration modellbasierter Systementwicklungsansätze in das Prozessmodell weiter ausgeformt. Neben Vorteilen beim Umgang mit vorhandener Komplexität beim Entwurf mechatronischer Systeme schildern die Autoren auch die Unterstützung der Anforderungsspezifikation durch frühe Modellierung der Systembeschreibung in semi"=formaler Notation wie der \foreignlanguage{english}{Systems Modeling Language} (SysML).

In einem Beitrag zum Entwurf autonomer Systeme trägt \textcite[Abschn.~2.3.2]{sifakis_system_2018} diverse Kritikpunkte am V"=Modell vor. Der Autor teilt die Auffassung, dass die im Modell abgebildete Annahme, dass initial alle Systemanforderungen in klar formulierter Form vorliegen, nicht realistisch ist. Ferner kritisiert \citeauthor{sifakis_system_2018} die Vorstellung eines konsequenten \foreignlanguage{english}{Top-Down}"=Systementwurfs, da in der Realität viel eher existierende Systeme inkrementell angepasst würden. Schließlich wird in dem Beitrag die im V"=Modell abgebildete Praxis einer Verifikation nach erfolgter Implementierung hinterfragt. Die Defizite führt der Autor als Grund an, dass sich die moderne Softwareentwicklung von dem V"=Modell zugunsten agiler Methoden abgewendet habe.

Bei der Vorstellung des V"=Modells als Referenzprozess für die Softwareentwicklung im \foreignlanguage{english}{Automotive}"=Bereich führen auch \textcite[Abschn.~1.5]{schauffele_automotive_2016} die bereits angesprochenen Einschränkungen des Prozessmodells als Abbildung des realen Entwicklungsgeschehens auf. Die Autoren betonen, dass in der Fahrzeugentwicklung zu Beginn Anforderungen der Benutzenden nicht vollständig bekannt seien, das V"=Modell aber implizit davon ausgehe. \citeauthor{schauffele_automotive_2016} teilen außerdem die Auffassung, dass in der Praxis eher inkrementelle und iterative Vorgehensweisen vorliegen, die die Phasen des V"=Modells mehrfach durchlaufen.

Obwohl deutlich wird, dass das V"=Modell für softwarelastige Entwicklungsgegenstände wie automatisierte Fahrzeuge keine faktische Prozessabbildung darstellt, hat es bis heute wesentliche Bedeutung zur Illustration des generellen Vorgehens und der Phasen einer Systementwicklung. Beispielsweise werden sowohl in der \foreignlanguage{english}{Automotive}"=Norm \textcite[Teil~1, Abbildung~1]{international_organization_for_standardization_iso_2018} als auch im Anhang der \textcite[Abbildung~A.15]{international_organization_for_standardization_isodis_2021} Prozessschritte in ein V"=Modell eingeordnet. In diversen ingenieurwissenschaftlichen Feldern existieren neben dem V"=Modell weitere etablierte Vorgehensmodelle für die Produktentwicklung. \textcite{eigner_uberblick_2014} bietet einen umfassenden Überblick über Geschichte und Gestalt diverser Vorgehensmodelle der Mechatronik und ihrer Teildisziplinen.

\subsection{Anforderungen an einen domänenspezifischen Referenzprozess}
\label{ssec42}
Die Vorstellung eines domänenspezifischen Referenzprozesses leistet einen wesentlichen Beitrag, die besonderen Anforderungen des automatisierten Fahrens an den Entwurf zu illustrieren. Hierfür sollen in diesem Abschnitt zunächst wichtige Anforderungen an ein spezifisches Prozessmodell beschrieben werden. Die Anforderungen werden im Folgenden durch verwandte Arbeiten und weiterführende Literatur gestützt ausführlich erläutert und am Ende dieses Abschnitts zusammengefasst aufgeführt.

\textcite[29]{daenzer_systems_1999} präsentieren vier Grundgedanken für Vorgehensmodelle der Systementwicklung. Die Autoren führen an, dass es zweckmäßig sei,
\begin{itemize}
\item vom Groben zum Detail vorzugehen,
\item alternative Lösungen zu erwägen,
\item die Systementwicklung in zeitlich unterscheidbare Prozessphasen zu teilen und
\item einen formalen Vorgehensleitfaden für die Lösung von Problemen anzuwenden.
\end{itemize}
\textcite[58-59]{daenzer_systems_1999} führen weiter aus, dass der Vorschlag zur Etablierung des Ansatzes \glqq{}vom Groben zum Detail\grqq{}~-- also eines \foreignlanguage{english}{Top-Down}"=Prinzips~-- eng mit weiteren Anforderungen an ein Vorgehensmodell verwandt ist. Einerseits kann das \foreignlanguage{english}{Top-Down}"=Vorgehen durch die Definition von Phasen als makroskopische Strategie zur Prozessstrukturierung konkretisiert werden. Andererseits ermöglicht eine strenge \foreignlanguage{english}{Top-Down}"=Systemspezifikation die ebenfalls empfohlene systematische Bildung und Evaluation von Lösungsvarianten.

\textcite{wilmsen_method_2020} definieren einen Prozess zur Anforderungsspezifikation an Referenzprozesse in großen Unternehmen. Die Au\-tor*innen legen notwendige Prozesscharakteristika anhand von Stakeholderanforderungen fest, wobei insbesondere verschiedene Nutzende des Prozesses wichtige Stakeholder darstellen (Ent\-wick\-ler*innen, Pro\-jekt\-ma\-na\-ger*innen etc.). \citeauthor{wilmsen_method_2020} wenden ihr Verfahren auf die \foreignlanguage{english}{Automotive}"=Vorentwicklung an und benennen einen hohen Detailgrad, eine Variabilität in den Arbeitsprodukten sowie das Ermöglichen einer agilen Arbeitsweise durch iterative und flexible Prozessmodelle als Beispiele wichtiger Anforderungen in diesem Feld.

In ihrer Arbeit greifen \textcite{wilmsen_method_2020} auch auf die Ergebnisse einer Befragung von Ex\-pert*innen aus großen deutschen Unternehmen der Automobilindustrie vorgestellt durch \textcite{pfeffer_automated_2019} zurück. Ausgehend von der Annahme, dass viele Entwicklungsprojekte konventioneller Fahrzeugsysteme bisher im Sinne eines V"=Modells durchgeführt werden, geben \citeauthor{pfeffer_automated_2019} Herausforderungen wieder, die befragte In\-dus\-trie\-ver\-tre\-ter*innen für eine Transformation der Prozessmodelle hin zur Entwicklung automatisierter Fahrfunktionen und Fahrzeuge identifiziert haben. Im Schnitt zeigte sich unter den Befragten deutliche Zustimmung, dass Entwicklungsprozesse und Prozessmodelle in der Automobilindustrie innerhalb der nächsten Dekade wesentliche Änderungen erfahren werden. Als häufig vorgetragene Erwartungen nennen die Au\-tor*innen dabei die Anpassung vorhandener Prozesse zur Einbindung agiler Methoden, um beispielsweise kürzere Entwicklungsschleifen insbesondere bei zugelieferten Teilsystemen zu etablieren.

Bereits in den vorangegangenen Abschnitten wurde die Wichtigkeit der systematischen Anforderungsdefinition an ein zu entwickelndes mechatronisches System deutlich. Automatisierte Fahrfunktionen und Fahrzeuge stellen komplexe Entwicklungsziele dar, deren Anforderungsanalyse die frühe Entwicklung bestimmt. Ziel einer Prozessbeschreibung sollte also auch die Berücksichtigung und konkrete Untergliederung dieser Entwicklungsphase sein. In Hinblick auf den Ablauf einer systematischen Definition notwendiger Sicherheitsziele leistet die Sicherheitsnorm \textcite{international_organization_for_standardization_iso_2018} bereits wichtige Beiträge (vgl.~\cite{winner_sicherheit_2024-1}). Der Sicherheitslebenszyklus der ISO~26262 beinhaltet eine Konzeptphase, die der Konkretisierung des zu entwickelnden Systems voransteht und implementierungsunabhängige Sicherheitsanforderungen spezifiziert. Dabei weist die Konzeptphase mehrere Teilschritte wie eine funktionale Spezifikation des Systems (Item"=Definition), eine Gefährdungsanalyse und eine Sicherheitskonzeptspezifikation auf. Allerdings wird diese Konzeptphase in dem in der ISO~26262 eingesetzten und auf dem V"=Modell basierenden Referenzprozess nicht abgebildet, sondern durch einen vorgelagerten Block beschrieben, der keine Rückschlüsse auf Struktur oder Phasen zulässt.

Domänenübergreifende Prozessbeschreibungen~-- wie in der VDI"=Richtlinie~2206~-- können an dieser Stelle auch keine konkrete Referenz darstellen, da diese die Anforderungen sicherheitskritischer Systeme nicht hinreichend berücksichtigen. Ein konkretes Beispiel ist die unabdingbar notwendige Durchführung einer Gefährdungsanalyse im Zuge des Entwurfs eines elektronischen Fahrzeugsystems. Obwohl die VDI"=Richtlinie~2206 die Entwicklung einer elektromechanischen Fahrzeugbremse als begleitendes Anwendungsbeispiel nutzt, wird in der Richtlinie nicht näher auf diesen Aspekt der Sicherheitsbetrachtung eingegangen.

\textcite{sexton_effective_2014} beschreiben in diesem Zusammenhang, dass während früher Entwicklungsaktivitäten eines elektronischen Fahrzeugsystems ein das V"=Modell begleitender informeller Prozess der Sicherheitsbetrachtung stattfindet. Den Prozess zeichne eine zyklische Verkettung von frühen Gefährdungs"= und Sicherheitsanalysen mit Designentscheidungen aus. Insbesondere beeinflussten Sicherheitsanforderungen aus potentiellen Gefährdungen die Systemarchitekturentscheidungen während der Konzeptphase fortlaufend (vgl. \autoref{ssec32}).

\textcite{reschka_fertigkeiten-_2017} greift für den Entwurf eines Fahrzeugführungssystems die in der ISO~26262 widergespiegelte Struktur des Entwicklungsprozesses auf und belegt damit eine Eignung der Prozessphasen der allgemeinen \foreignlanguage{english}{Automotive}"=Sicherheitsnorm für den Anwendungsfall des automatisierten Fahrens. Daneben schlägt \citeauthor{reschka_fertigkeiten-_2017} aber insbesondere auch eine Erweiterung der Vorgaben für die Konzeptphase in Form eines strukturierten Prozesses zur systematischen Erzeugung einer Item-Definition für automatisierte Fahrfunktionen vor.

Zusammenfassend sollte ein domänenspezifisches Prozessmodell für den Entwurf automatisierter Fahrfunktionen und Fahrzeuge also
\begin{itemize}
\item Prozessphasen unterscheiden,
\item ein \foreignlanguage{english}{Top-Down}"=Prinzip etablieren,
\item den Vergleich unterschiedlicher Lösungen ermöglichen,
\item den agilen Charakter der frühen Entwicklung abbilden,
\item die Prozessphase der Anforderungsdefinition abbilden und
\item den Stand der Technik berücksichtigen (zum Beispiel den Referenzprozess der ISO~26262).
\end{itemize}

\subsection{Systematischer Entwurf automatisierter Fahrfunktionen und Fahrzeuge}
\label{ssec43}
Im Folgenden wird ein Referenzprozess vorgestellt, der einen systematischen Entwurf im Bereich des automatisierten Fahrens beschreibt und die zuvor aufgeführten Anforderungen an eine Prozessbeschreibung berücksichtigt. Das vorgestellte Modell basiert auf der Prozessbeschreibung der Vorentwicklung einer automatischen Notbremse als Assistenzsystem für Straßenfahrzeuge~\cite{maurer_unfallschwereminderung_2002}. Eine ausführliche Beschreibung der Hintergründe des Entwicklungsprojekts erfolgt durch \textcite{winner_entwicklungsprozess_2015} in einer früheren Version dieses Handbuchs. Darin wird unter anderem der besondere Nutzen einer Frontkollisionsvermeidung durch Assistenzsysteme aufgezeigt, aber auch die wegbereitende technologische Entwicklung bei Wahrnehmungssystemen sowie die weiterhin vorhandene Unsicherheit für Eingriffsentscheidungen erläutert. Außerdem wird näher auf die Herausforderung eines belastbaren Sicherheitskonzepts für Assistenzsysteme insbesondere auch unter ergonomischen Aspekten eingegangen.

\begin{figure}
        \centering
        \includegraphics[width=\textwidth]{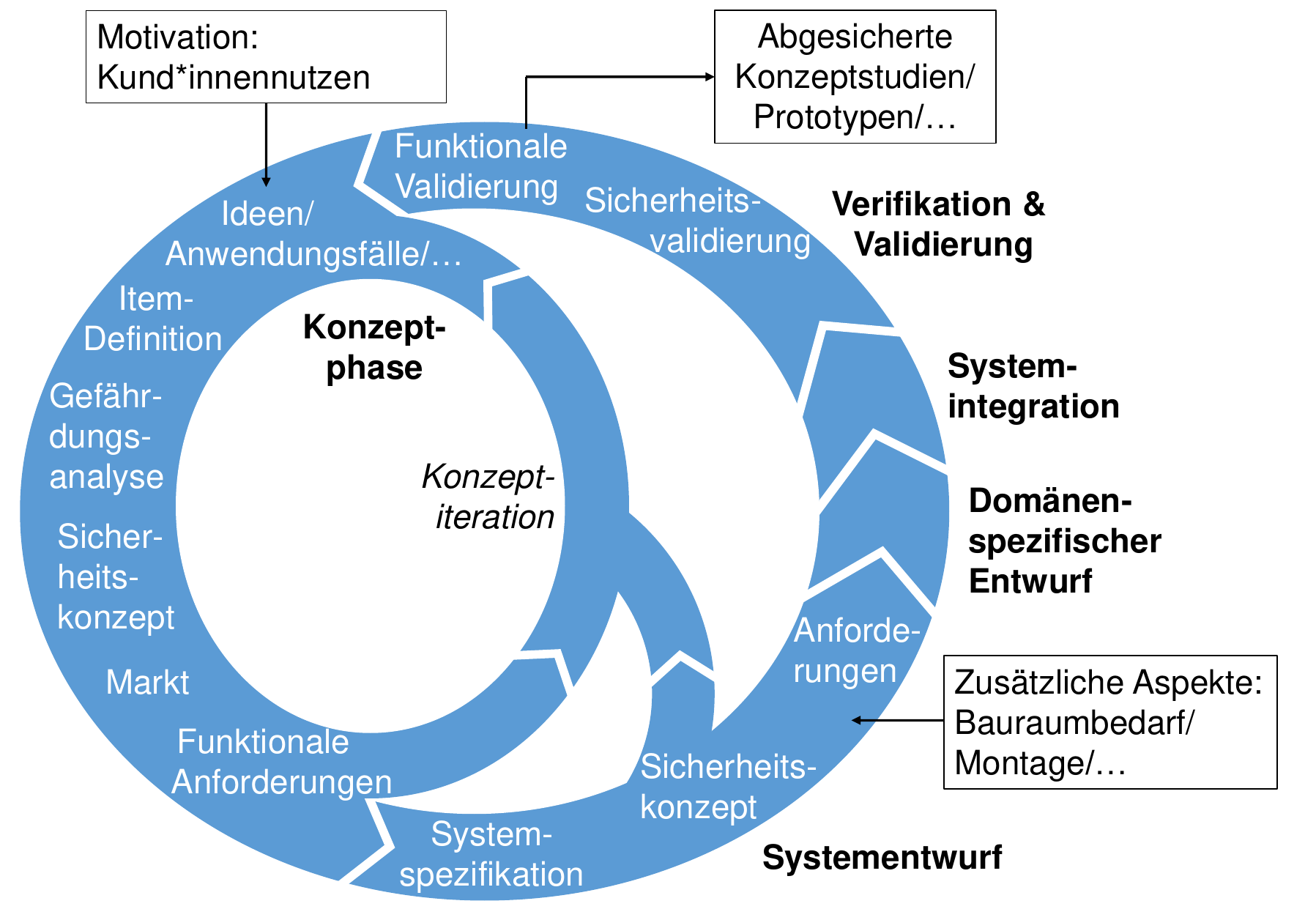}
        \caption{Systematischer Entwurf automatisierter Fahrfunktionen und Fahrzeuge nach~\cite{graubohm_systematic_2017}}
        \label{fig2}
        \end{figure}

\autoref{fig2} zeigt den Referenzprozess für einen systematischen Entwurf automatisierter Fahrfunktionen und Fahrzeuge, der auf dieser Basis entwickelt wurde. In seiner hier vorgestellten Form wurde der Referenzprozess erstmals in~\cite{graubohm_systematic_2017} erläutert. Die Anpassung des Prozessmodells zur Abbildung der Herausforderungen des automatisierten Fahrens beruht auf Erkenntnissen aus dem Forschungsprojekt aFAS, das in \autoref{sec5} als Anwendungsbeispiel vorgestellt wird.

Das Prozessmodell weist eine zyklische Form mit zwei möglichen Schleifen auf und soll dadurch insbesondere den iterativen Charakter des Entwurfs betonen. Die Iteration über die innere Schleife wird als \emph{Konzeptiteration} bezeichnet und meint eine besonders ressourcenschonende Revision der Funktionsspezifikation in einem frühen Entwicklungsstadium. Die Teilschritte der inneren Schleife sind gekennzeichnet durch Diskussionen über Nutzen, Realisierbarkeit und Konsequenzen angestrebter Funktionsumfänge mit Ex\-pert*innen verschiedener Fachbereiche. Das Auslösen einer inneren Iteration ist dabei zu jedem Zeitpunkt innerhalb der ersten Entwicklungsphasen möglich, sobald unlösbare Konflikte aufgrund von Entwurfsentscheidungen auftreten. Für die erste illustrierte Entwicklungsphase der systematischen Definition funktionaler technologieunabhängiger Anforderungen wird in Anlehnung an die \foreignlanguage{english}{Automotive}"=Norm ISO~26262 die Bezeichnung \emph{Konzeptphase} gewählt. Die weiteren Bezeichnungen \emph{Systementwurf}, \emph{domänenspezifischer Entwurf}, \emph{Systemintegration} sowie \emph{Verifikation \& Validierung} orientieren sich an den Phasen des Referenzprozesses der VDI"=Richtlinie~2206 (vgl. \autoref{ssec41}).

Konkret beschreibt das Prozessmodell den Beginn der Konzeptphase als Sammlung und Dokumentation von Ideen, Anwendungsfällen etc., die aus der Motivation resultieren, einen möglichst umfangreichen Kund*innen\-nut\-zen anzubieten. Der zentrale Nutzen eines automatisierten Fahrzeugs ist das Angebot von Mobilität, daneben werden Nutzende von weiteren Aspekten wie Komfort, Effizienz oder Sicherheit profitieren. Erfasste Ideen und Anwendungsfälle stellen im weiteren Verlauf des Entwurfs eine zentrale Zielvorgabe dar. Einschränkungen des Funktionsumfangs können später aus unterschiedlichen Gründen beschlossen werden und erfordern häufig eine Anpassung der angestrebten Anwendungsfälle. Das wird im Prozessmodell durch die zyklische Gestalt und in beiden Schleifen ausgedrückt.

Im Anschluss an die Sammlung von Anwendungsfällen wird die angestrebte Funktionsspezifikation abstrakt in einer Item"=Definition erfasst, die auch Aspekte wie die vorläufige Struktur und erwartete Einsatzumgebungen des zu entwickelnden Systems adressiert. Dies stellt die Basis einer anschließenden Gefährdungsanalyse dar. Darin werden~-- ausgehend von möglichen Betriebssituationen~-- Gefährdungen durch Fahrzeugverhalten identifiziert und eine Risikobewertung durchgeführt. Anschließend muss ein für den angestrebten Funktionsumfang geeignetes Sicherheitskonzept entwickelt werden. Im Zuge der Konzeptphase wird ein funktionales Sicherheitskonzept erstellt, das Sicherheitsmechanismen und Sicherheitsstrategien zur Reduzierung identifizierter Risiken auf ein akzeptables Maß implementierungsunabhängig durch funktionale Sicherheitsanforderungen beschreibt (vgl.~\cite[Abschn.~6.2.2.2]{winner_sicherheit_2024-1}).

Die im Prozessmodell abgebildete Relevanz früher Sicherheitsanalysen folgt dem Prinzip des \emph{\foreignlanguage{english}{Safety-by-design}}: Gefährdungen, die aus einem angestrebten Funktionsumfang und Einsatzkontext resultieren, werden frühzeitig im Zuge des Entwurfs identifiziert und aktiv durch Anpassung eines vorläufigen Systemdesigns oder Konzipierung und Implementierung geeigneter Mechanismen möglichst umfassend vermieden. \foreignlanguage{english}{Safety-by-design} steht somit im Gegensatz zu Entwurfsansätzen, die aufgrund guter fachlicher Praxis während der Entwicklung von einer abschließend durch Erprobung eines Systems feststellbaren ausreichenden Sicherheit ausgehen. Der technische Bericht \textcite{international_organization_for_standardization_isotr_2020} macht deutlich, dass eine konsequente Anwendung des \foreignlanguage{english}{Safety-by-design}"=Prinzips entscheidend zur Sicherheit automatisierter Fahrzeuge beiträgt.

Der systematische Entwurf bildet die in \autoref{ssec32} angeführte Bedingung ab, dass eine Entwicklung bestimmter Funktionsausprägungen im Bereich des automatisierten Fahrens nur dann weiterverfolgt wird, wenn die Umsetzung eines systemweiten Sicherheitskonzepts, das potentielle Gefährdungen hinreichend adressiert, möglich scheint. Andernfalls muss die Funktionsspezifikation angepasst werden, sodass umsetzbare Sicherheitsstrategien ein belastbares Sicherheitskonzept darstellen. Zum Beispiel kann die Einsatzumgebung eingeschränkt oder auf Anwendungsfälle und damit Funktionsumfänge verzichtet werden. Im Prozessmodell ist dies ein Hauptaspekt der \emph{Konzeptiteration}: Wenn kein zufriedenstellendes Sicherheitskonzept vorliegt, sollte die funktionale Zielsetzung angepasst werden, anstatt unnötig Ressourcen in die weitere Umsetzung von Konzepten zu stecken, die als nicht absicherbar angenommen werden müssen. Eine Entsprechung dieses Mechanismus innerhalb des Entwurfs von Automotive-Funktionen findet sich auch in der Sicherheitsnorm ISO/""DIS~21448. Die Norm schlägt eine Anpassung der funktionalen Systemspezifikation oder Einschränkung der Anwendungsfälle im Zuge der Entwicklung und als Teil der Sicherheitsanalyse vor, wenn Risiken oberhalb eines akzeptablen Niveaus festgestellt werden, die nicht weiter reduziert werden können~\cite[Abschn.~8]{international_organization_for_standardization_isodis_2021}.

Unter dem anschließenden Stichpunkt \emph{Markt} regt das Prozessmodell zu einer kritischen Überprüfung an, ob die angestrebten Funktionsumfänge~-- insbesondere auch unter dem Aspekt potentieller Einschränkungen durch das Sicherheitskonzept~-- ein marktfähiges Konzept darstellen, dessen weitere Konkretisierung und Umsetzung den Entwickelnden vorteilhaft erscheint. Bis zu diesem Punkt sind nur begrenzte Ressourcen in vornehmlich konzeptuelle Arbeiten eingeflossen. Die tatsächliche Erzeugung von Konzeptstudien und Prototypen sowie deren Evaluierung in umfangreichen Tests erfordert ein Vielfaches der Zeit, monetären sowie anderer Ressourcen und sollte nur für vielversprechende Konzepte erfolgen.

Als Produkt der Konzeptphase wird ein belastbares Konzept durch die Spezifikation funktionaler Anforderungen ausgedrückt. Neben der Spezifikation angestrebter Funktionsumfänge fließen an dieser Stelle auch sämtliche Sicherheitsanforderungen an funktionale Komponenten aus dem zuvor konzipierten Sicherheitskonzept ein. Wie schon im \autoref{sec2} erläutert sind die funktionalen Anforderungen möglichst lösungsneutral~-- also technologieunabhängig und implementierungsunspezifisch~-- zu formulieren, um den Lösungsraum nicht unnötig einzuschränken.

Auf Basis der funktionalen Anforderungen erfolgt dann die Systemspezifikation im Zuge der Prozessphase Systementwurf. Hierbei werden erstmals konkrete Technologieentscheidungen notwendig, die in eine konkret in Hardware und Software zu implementierende technische Systemarchitektur einfließen (vgl.~\cite{winner_architektursichten_2024}). Die Auswahl, welche Steuergeräte, Sensoren und Aktuatoren eingesetzt und wie diese miteinander verschaltet werden, stellt dann eine Grundlage für die weitere Konkretisierung des Sicherheitskonzepts dar. Für automatisierter Fahrfunktionen und Fahrzeuge ist die Festlegung eines sogenannten technischen Sicherheitskonzepts, das die notwendige Umsetzung funktionaler Sicherheitsanforderungen in konkreten mechatronischen Komponenten spezifiziert (vgl.~\cite{winner_sicherheit_2024-1}), ein zentraler und gleichzeitig sehr herausfordernder Prozessschritt des Systementwurfs.

Das Prozessmodell sieht an dieser Stelle eine erneute Möglichkeit der Konzeptiteration vor. Diese Möglichkeit einer späten Konzeptanpassung wird insbesondere dadurch erforderlich, dass erst an dieser Stelle ein Abgleich verfügbarer technologischer Leistungsfähigkeit mit den funktionalen Anforderungen erfolgt. Ein einfaches Beispiel hierfür wären Anforderungen, die eine Leistungsfähigkeit von der Umfeldsensorik (zum Beispiel hinsichtlich Erfassungsbereich oder Reichweite) verlangen, die durch keine bekannte Sensortechnologie darstellbar ist. An dieser Stelle würde es keinen Sinn ergeben, die Erzeugung einer Konzeptstudie weiterzuverfolgen, sondern es muss eine Anpassung der Anforderungen auf Basis eines erneuten Durchlaufens der Konzeptphase erfolgen.

Den Abschluss der Prozessphase Systementwurf stellt im Modell die Spezifikation technischer Anforderungen dar. Diese Anforderungen konkretisieren die vorher als funktionale Anforderungen vorliegende Konzeptbeschreibung auf Basis der durchgeführten technischen Spezifikation und anschließenden Sicherheitsbetrachtung. Außerdem fließen weitere~-- gegebenenfalls \foreignlanguage{english}{Automotive}"=spezifische~-- Aspekte wie Bauraumbedarf, Montage etc. ein. Die Anforderungsliste stellt den Ausgangspunkt einer domänenspezifischen Umsetzung des Systems in Hardware und Software dar und muss entsprechend belastbar und umfassend sein.

Die Implementierung der mechatronischen Teilsysteme durch spezialisierte Entwickelnde erfolgt anschließend in eigenen Prozessen auf Basis der technischen Systembeschreibung. Dabei wird im Regelfall eine parallele Umsetzung einzelner Komponenten erfolgen, die eine anschließende Integration zu einem Gesamtsystem erforderlich macht. Im Anschluss an die Erzeugung und Integration sämtlicher Hardware"= und Softwarekomponenten des entwickelten Systems können Systemtests im Zuge der Verifikation und Validierung erfolgen. Hierbei unterscheidet das Prozessmodell insbesondere die Validierung des systemweiten Sicherheitskonzepts und die funktionale Validierung.

Im Zuge der Sicherheitsvalidierung wird überprüft, ob das Sicherheitskonzept vorstellungsgemäß umgesetzt und seine Wirkung im Sinne einer Gefährdungsvermeidung im vorgesehenen Einsatzkontext gemäß Tests ausreichend scheint. Die funktionale Validierung beschreibt hingegen empirische Maßnahmen zur Überprüfung, ob die erzeugte Konzeptstudie beziehungsweise der Prototyp den initial als Motivation dokumentierten Kund*innen\-nut\-zen ausreichend bedient. Erkenntnisse der Systemtests und insbesondere identifizierte Defizite des in der letzten Iteration erzeugten Systems dienen als wichtige Informationsquelle für eine nächste Iteration im Prozessmodell, um schließlich gemäß der VDI~2206 Definition des Entwurfs (vgl. \autoref{sec2}) ein abgesichertes Konzept für ein potentielles Serienprodukt zu erzeugen.

Grundsätzlich ist festzustellen, dass die zyklische Gestalt des vorgestellten Prozessmodells die mögliche Wiederverwendung bereits erzeugter Artefakte sowohl im Zuge der inneren als auch der äußeren Iteration über das Endprodukt des letzten Durchlaufs hinaus betont. Konkret können große Teile der Sicherheitsanalyse, konzipierter Wirkmechanismen und Systemarchitekturen, funktionaler Anforderungen etc. auch über mehrere Iterationen einer Konzeptstudie wiederverwendet oder nur geringfügig revidiert und erweitert werden. Diesen Aspekt teilt sich der beschriebene systematische Entwurf automatisierter Fahrfunktionen und Fahrzeuge mit dem Spiralmodell der Softwareentwicklung nach \textcite{boehm_spiral_1988}. Das vorgestellte Prozessmodell ergänzt aber die Aspekte einer kontinuierlichen Anforderungsrevision und Rückführung zur initialen Zielformulierung insbesondere über den Mechanismus der Konzeptiteration. Grundsätzlich bedient der hochgradig iterative Charakter des vorgestellten Modells außerdem das in agilen Entwicklungsmethoden abgebildete Prinzip, ein Produkt inkrementell in kurzen Iterationen zu erzeugen und sich dadurch an kontinuierliche Veränderungen der Randbedingungen eines Entwicklungsprojekts anpassen zu können~\cite{beck_manifesto_2001}.

Nach \textcite{cooper_process_1983} handelt es sich bei dem hier erläuterten Modell des systematischen Entwurfs um ein deskriptives Prozessmodell, in dem Prozessbeobachtungen aus der Praxis im Sinne einer Referenz abgebildet werden. Entsprechend ist der illustrierte systematische Entwurf automatisierter Fahrfunktionen und Fahrzeuge (vgl. \autoref{fig2}) kein normatives Modell mit dem Ziel, einen Idealprozess widerzuspiegeln. Zwar teilen viele normative Modelle den Aspekt der Definition unterscheidbarer Prozessphasen; die beliebig vielen Iterationen bis zur Erzeugung eines abgesicherten Konzepts im Zuge des Entwurfs stellen aber keinen idealen Leitfaden dar. Hier müsste wahrscheinlich ein Prozess ohne Iterationen angestrebt werden, der aber in der Praxis nicht realistisch ist.

Wie bereits in \autoref{sec3} erläutert ist die Entwicklung automatisierter Fahrfunktionen und Fahrzeuge von einer Vielzahl potentiell konfliktärer Stakeholderinteressen geprägt. Die Sicherheit entwickelter Systeme ist hierbei ein zentraler Aspekt und führt zu einer Konzentration des vorgestellten Prozessmodells auf Sicherheitsaspekte schon in der frühen Entwurfsphase. Die gesellschaftliche Toleranz gegenüber einem realisierbaren Sicherheitsniveau muss allerdings~-- auch durch explizite Offenlegung verbleibender Residualrisiken~-- entwurfsbegleitend überprüft werden. Eine systematische Identifikation und Berücksichtigung der Stakeholdergruppen ist geeignet, weitere wichtige, über Sicherheitsaspekte hinausgehende Randbedingungen für die Konzeptspezifikation zu ermitteln. Hierfür können Methoden, die der Berücksichtigung menschlicher Werte im Zuge der Systementwicklung dienen, wichtige Beiträge darstellen. Ein Ansatz, das werteorientierte Entwicklungsparadigma \emph{\foreignlanguage{english}{Value-Sensitive-Design}} nach \textcite{doorn_value_2013} mit der Prozessstruktur aus \autoref{fig2} zu verbinden, wird in~\cite{graubohm_value_2020} vorgestellt.

\section{Anwendungsfall aFAS}
\label{sec5}
Das 2018 erfolgreich mit einer öffentlichen Demonstration im fließenden Verkehr abgeschlossene Forschungsprojekt aFAS (automatisch fahrer*innen\-los fahrendes Absicherungsfahrzeug für Arbeitsstellen auf Bundesautobahnen) diente als wichtige Informationsbasis zur Weiterentwicklung des ursprünglich für den Entwurf von Assistenzsystemen konzipierten Referenzprozesses. Das aktuelle Prozessmodell wurde in \autoref{ssec43} vorgestellt und soll an dieser Stelle am Beispiel der im aFAS Projekt durchgeführten Entwicklung eines Prototyps eines fahrer*innen\-lo\-sen Absicherungsfahrzeuges angewandt werden.

Wie im Prozessmodell beschrieben war auch im aFAS Projekt die Motivation der Entwicklung eines automatisierten Absicherungsfahrzeugs ein bestimmter Kund*innen\-nut\-zen: die Sicherheit der Beschäftigten. Für menschliche Fahrende eines eine Arbeitsstelle ankündigenden Absicherungsfahrzeugs besteht eine reelle Gefahr lebensbedrohlicher Verletzungen durch Unfälle wegen Unachtsamkeit der Vorbeifahrenden. Im Forschungsprojekt sollte deshalb ein System entwickelt werden, das die langsame Fahrt auf den Seitenstreifen von Autobahnen automatisiert und damit den fahrer*innen\-lo\-sen Betrieb im öffentlichen Straßenverkehr möglich macht.

\begin{figure}
        \centering
        \includegraphics[width=.49\textwidth]{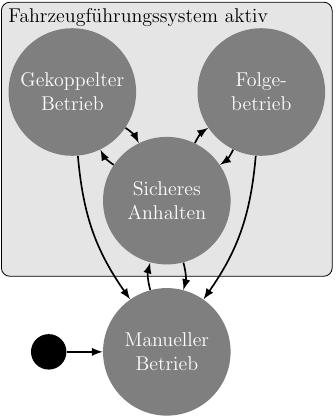}
        \caption{Betriebsmodi des automatisierten Absicherungsfahrzeugs im Projekt aFAS nach~\cite{stolte_towards_2015}}
        \label{fig3}
        \end{figure}

\begin{figure}
        \centering
        \includegraphics[width=.69\textwidth]{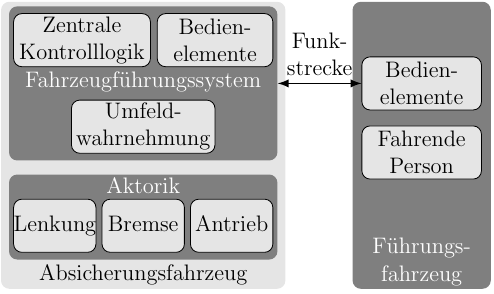}
        \caption{Funktionale Komponenten des Gesamtsystems im Projekt aFAS nach~\cite{graubohm_functional_2019}}
        \label{fig4}
        \end{figure}

Auf Basis dieser Zielsetzung wurden unterschiedliche Anwendungsfälle konkretisiert, wie eine reguläre manuelle Operation des Absicherungsfahrzeugs für die Fahrt zwischen Arbeitsstellen und dem Betriebshof ohne Systemeingriffe und automatisierte Folgefahrten mit einem Arbeitsfahrzeug als Führungsfahrzeug in unterschiedlichen Distanzen. Die angestrebten Funktionsumfänge wurden anschließend gemeinsam mit vorläufigen funktionalen Systemstrukturannahmen und weiteren Randbedingungen in einer Item"=Definition dokumentiert. \textcite{ohl_autonomes_2012} sowie \textcite{stolte_towards_2015} erläutern wesentliche Inhalte dieses im Laufe eines Entwurfs lebendigen Dokuments. An dieser Stelle werden Illustrationen der Betriebsmodi (\autoref{fig3}) und der funktionalen Systemstruktur (\autoref{fig4}) aufgeführt, die als zentrale Elemente der Funktionsspezifikation in der Item"=Definition und für die weitere Funktionsentwicklung dienten.

Mit den in einer Item"=Definition dokumentierten Annahmen über Funktionsumfänge, Systemstruktur und den Einsatzrahmen konnte im Zuge des Forschungsprojekts die Bearbeitung der nächsten im Prozessmodell beschriebenen Aufgabe beginnen: die Durchführung einer Gefährdungsanalyse. Struktur, Inhalte und Ergebnisse der Gefährdungsanalyse für das automatisierte Absicherungsfahrzeug werden durch \textcite{stolte_hazard_2017} beschrieben. Zentrale Bedeutung hat hierbei, dass die im Kontext eines fahrer*innen\-lo\-sen Betriebs zu untersuchenden Ursachen für Gefährdungen nicht ausschließlich in der Möglichkeit von Fehlern und Ausfällen elektronischer und elektrischer Systemkomponenten, die den Fokus der Sicherheitsnorm ISO~26262 darstellt (vgl.~\cite{winner_sicherheit_2024-1}), liegen. Vielmehr wurden innerhalb der Sicherheitsanalyse des Entwicklungsprojekts auch ergonomische Aspekte~-- wie zu Beispiel mögliche Unklarheit für Bedienende über den aktuellen Betriebszustand des Fahrzeugs~-- untersucht. Darüber hinaus wurden funktionale Unzulänglichkeiten des Systems als mögliche Gefährdungsursachen evaluiert. Als Beispiele sind hier die unzureichende Erkennung der Fahrstreifenmarkierung oder des Führungsfahrzeugs durch das Fahrzeugführungssystem anzuführen. Insofern wurde in der Gefährdungsidentifikation ein Vorgehen realisiert, das inzwischen in der Norm ISO/""DIS~21448 beschrieben und gefordert ist.

Eine Gefährdungsanalyse wird prinzipiell durch die Spezifikation von Sicherheitszielen, die sämtliche identifizierten Risiken adressieren, vorerst abgeschlossen. Bevor es im Zuge der Gefährdungsanalyse für das automatisierte Absicherungsfahrzeug dazu kommen konnte, wurde eine Konzeptiteration, die im Prozessmodell durch die innere Schleife dargestellt ist, notwendig. \textcite{stolte_hazard_2017} beschreiben, dass ursprünglich die Realisierung eines Anwendungsfalls vorgesehen war, der das Passieren eines stationären Hindernisses auf dem Seitenstreifen der Autobahn durch Eintritt des automatisierten Absicherungsfahrzeugs in den rechten Fahrstreifen beinhaltet. Im Zuge der Sicherheitsbetrachtungen wurde der große Umfang des Gefährdungspotentials einer solchen Manöversequenz in fließendem Verkehr festgestellt. Mechanismen zur Abwendung der resultierenden Risiken waren im Projektzeitraum nicht realisierbar. Insbesondere wäre es durch diesen Sonderfall unmöglich geworden, dem automatischen Fahrzeugführungssystem im Sicherheitskonzept generell zu untersagen, die Seitenstreifenmarkierung zu überfahren und in den rechten Fahrstreifen einzutreten. Insofern lag ein wesentlicher Auslegungskonflikt vor, der durch eine Iteration innerhalb der Konzeptphase aufgelöst werden musste. Im Zuge dieser Iteration wurde der besonders herausfordernde Anwendungsfall vorerst ausgeschlossen und anschließend die Item"=Definition entsprechend angepasst. Mit den angepassten Funktionsumfängen war dann ein Abschluss der Gefährdungsanalyse und Risikobewertung durch Spezifikation von 17 Sicherheitszielen mit entsprechend erforderlichen Integritätsniveaus (s.~\cite{stolte_hazard_2017}) möglich.

Wie im Prozessmodell illustriert war der nächste Schritt des Entwurfs im Forschungsprojekt die Erstellung eines funktionalen Sicherheitskonzepts. Hierfür wurden die Sicherheitsziele aufgegriffen und durch eine Erarbeitung von Sicherheitsmechanismen und Sicherheitsstrategien konkretisiert. Aus den definierten Zielen wurden dann (möglichst lösungsunabhängige) funktionale Sicherheitsanforderungen für einzelne der in \autoref{fig4} widergespiegelten Komponenten spezifiziert. Tatsächlich flossen hierbei aber auch einige bereits im Projekt getroffene Entscheidungen über die Struktur und Kompetenzen einzelner Steuergeräte ein. Die konkreten Inhalte des im Projekt aFAS erstellten Sicherheitskonzepts werden in~\cite{graubohm_functional_2019} erläutert.

Aufgrund seines Charakters als Forschungsprojekt waren Marktanalysen, die im Prozessmodell als nächste Instanz aufgeführt werden, nicht von großer Bedeutung. Die Realisierung des Automatisierungssystems mit begrenzten Projektressourcen und die erfolgreiche Demonstration ohne Anwesenheit einer Person im Absicherungsfahrzeug erhielt jedoch im Anschluss an den Projektabschluss einige Aufmerksamkeit, da durch das Projekt eine Lösung eines national und international relevanten Sicherheitsproblems präsentiert wurde~\cite{jentzsch_man_2018}.

Das Zusammenstellen der funktionalen Anforderungen, die das Sicherheitskonzept, die Funktionsumfänge, Systemstruktur etc. beschreiben, bildete auch im Projekt aFAS den Abschluss der Konzeptphase. Das Systemdesign wurde anschließend durch beteiligte Industriepartner konkretisiert und im Zuge dessen das Sicherheitskonzept in technische Anforderungen überführt. Gleiches gilt auch für Umsetzung und Integration der einzelnen Systemkomponenten.

Die für eine Demonstration des Prototyps erforderlichen Tests zur Verifikation und Validierung der Funktion und Sicherheit wurden abschließend in unterschiedlichen Umgebungen durchgeführt. Auch wurden gezielt Fehler injiziert, um die Wirkung von Sicherheitsmechanismen zu untersuchen~\cite{bagschik_funktionale_2018}. Zunächst fuhr das Absicherungsfahrzeug durch das automatisierte Fahrzeugführungssystem kontrolliert auf abgesperrter Teststrecke, dann auf einem unbefahrenen Autobahnneubau und schließlich im fließenden Verkehr auf dem für die Demonstration vorgesehenen Autobahnabschnitt. Fahrzeugführende des Absicherungsfahrzeugs hatten während der Tests Möglichkeiten, in die Fahrzeugführung einzugreifen. Während der eigentlichen Abschlussdemonstration befand sich die Person aber nicht mehr im Absicherungsfahrzeug, sondern im Führungsfahrzeug und hatte lediglich die Möglichkeit, das Absicherungsfahrzeug per Funkbefehl anzuhalten.

\section{Zusammenfassung}
\label{sec6}
Die Anwendung geeigneter Methoden und Prozesse leistet bereits in frühen Phasen einen wesentlichen Beitrag zum Erfolg einer Entwicklung automatisierter Fahrfunktionen und Fahrzeuge. In diesem Kapitel wurden die Randbedingungen eines Entwurfs im Kontext des automatisierten Fahrens erläutert. Dazu wurde zunächst der Entwurfsbegriff innerhalb der Entwicklung komplexer mechatronischer Systeme behandelt und anschließend auf die besonderen Herausforderungen bei der Durchführung eines Entwurfs von automatisierten Fahrfunktionen und Fahrzeugen eingegangen. Die Überwindung diverser identifizierter Herausforderungen kann durch eine Strukturierung des Entwurfsprozesses unterstützt werden. Entsprechend wurde ein Prozessmodell für den systematischen Entwurf im automatisierten Fahren vorgestellt und erläutert. Neben einer Referenz für unterscheidbare Phasen etabliert der Referenzprozess auch Mechanismen, die unnötige Ineffizienzen durch anhaltendes Hinterfragen der Funktionsspezifikation und weiterer Entwicklungsentscheidungen vermeiden. Abschließend wurde über die erfolgreiche Durchführung des Entwurfs eines automatisierten Fahrzeugführungssystems für ein Absicherungsfahrzeug aus dem Forschungsprojekt aFAS berichtet, die wesentliche Beiträge zur aktuellen Version des Prozessmodells leistete.

\section*{Danksagung}
Dieses Kapitel verarbeitet Ergebnisse des Forschungsprojekts aFAS. Das Projekt aFAS wurde unterstützt durch das Bundesministerium für Wirtschaft und Energie (BMWi). Das Projektkonsortium bestand aus MAN (Konsortialführer), ZF TRW, WABCO, Bosch Automotive Steering, Technische Universität Braunschweig, Hochschule Karlsruhe, Hessen Mobil~-- Straßen"= und Verkehrsmanagement und BASt~-- Bundesanstalt für Straßenwesen. Die Autoren danken ihren aktuellen oder ehemaligen Kollegen \mbox{Andreas} \mbox{Reschka}, \mbox{Gerrit} \mbox{Bagschik}, \mbox{Torben} \mbox{Stolte} und \mbox{Markus} \mbox{Steimle} für ihre Arbeit an der Entwicklung eines automatisierten Absicherungsfahrzeugs im Projekt aFAS im Zuge ihrer Beschäftigung am Institut. Ihre Beiträge stellen die Grundlage der hier erläuterten Weiterentwicklung des Prozessmodells für den systematischen Entwurf im automatisierten Fahren dar.

Die Forschungsarbeiten zum Entwurf automatisierter Fahrfunktionen und Fahrzeuge wurden teilweise im Rahmen des Projekts \glqq{}UNICAR\textit{agil}\grqq{} durchgeführt (FKZ~16EMO0285). Wir bedanken uns für die finanzielle Unterstützung des Projekts durch das Bundesministerium für Bildung und Forschung (BMBF).

\printbibliography

@techreport{sae_international_sae_2021,
	type = {{SAE International} \foreignlanguage{ngerman}{Richtlinie}},
	title = {{Taxonomy} and {Definitions} for {Terms} {Related} to {Driving} {Automation} {Systems} for {On}-{Road} {Motor} {Vehicles}},
	number = {{SAE}~J3016\_202104},
	author = {{SAE~J3016}},
	year = {2021},
	langid= {english},
doi = {10.4271/J3016_202104},
}

@incollection{winner_testkonzepte_2024,
	address = {Wiesbaden},
	title = {Testkonzepte f{\"u}r die {Absicherung} von automatisiertem {Fahren}},
	isbn = {978-3-658-38485-2 978-3-658-38486-9},
	url = {https://link.springer.com/10.1007/978-3-658-38486-9_46},
	language = {de},
	urldate = {2025-01-06},
	booktitle = {Handbuch {Assistiertes} und {Automatisiertes} {Fahren}},
	publisher = {Springer},
	author = {Eckstein, Lutz and Winner, Hermann},
	editor = {Winner, Hermann and Dietmayer, Klaus and Eckstein, Lutz and Jipp, Meike and Maurer, Markus and Stiller, Christoph},
	year = {2024},
	doi = {10.1007/978-3-658-38486-9_46},
	series = {ATZ/\allowbreak{}MTZ-Fachbuch},
	pages = {1253--1287},
	edition = {4.~Aufl},
}

@inproceedings{maurer_hochautomatisiertes_2018,
	address = {Goslar},
	title = {Hochautomatisiertes und vollautomatisiertes {Fahren}},
	isbn = {3-472-09576-8},
	booktitle = {56.~{Deutscher} {Verkehrsgerichtstag} 2018},
	publisher = {Luchterhand},
	author = {Maurer, Markus},
	year = {2018},
	pages = {43--57},
}

@book{suh_axiomatic_2001,
	address = {New York, NY, USA},
	series = {The {Oxford} series on advanced manufacturing},
	title = {Axiomatic design: advances and applications},
	isbn = {0-19-513466-4},
	shorttitle = {Axiomatic design},
	language = {eng},
	publisher = {Oxford University Press},
	author = {Suh, Nam P.},
	year = {2001},
	langid= {english},
}

@techreport{verein_deutscher_ingenieure_entwicklungsmethodik_2004,
	type = {{Verein Deutscher Ingenieure} {Richtlinie}},
	title = {Entwicklungsmethodik f{\"u}r mechatronische {Systeme}},
	number = {{VDI}~2206:2004-06},
	author = {{VDI~2206}},
	year = {2004},
}

@book{daenzer_systems_1999,
	address = {Z{\"u}rich},
	edition = {10.,~durchgesehene Aufl},
	title = {\foreignlanguage{english}{Systems {Engineering}}~{\textendash} {Methodik} und {Praxis}},
	isbn = {3-85743-998-X},
	language = {de},
	publisher = {Industrielle Organisation},
	author = {Haberfellner, Reinhard and Nagel, Peter and Becker, Mario and B{\"u}chel, Alfred and {von Massow}, Heinrich},
	editor = {Daenzer, Walter F. and Huber, Fritz},
	year = {1999},
}

@incollection{winner_sicherheit_2024-1,
	address = {Wiesbaden},
	title = {Sicherheit von assistierten und automatisierten {Systemen}},
	isbn = {978-3-658-38485-2 978-3-658-38486-9},
	url = {https://link.springer.com/10.1007/978-3-658-38486-9_6},
	language = {de},
	urldate = {2025-01-31},
	booktitle = {Handbuch {Assistiertes} und {Automatisiertes} {Fahren}},
	publisher = {Springer},
	author = {Wilhelm, Ulf and Ebel, Susanne},
	editor = {Winner, Hermann and Dietmayer, Klaus and Eckstein, Lutz and Jipp, Meike and Maurer, Markus and Stiller, Christoph},
	year = {2024},
	doi = {10.1007/978-3-658-38486-9_6},
	series = {ATZ/\allowbreak{}MTZ-Fachbuch},
	pages = {91--127},
	edition = {4.~Aufl},
}

@techreport{international_organization_for_standardization_iso_2018,
	type = {{International Organization for Standardization} \foreignlanguage{ngerman}{Norm}},
	title = {{Road} vehicles~{\textemdash} {Functional} safety},
	number = {{ISO}~26262:2018},
	author = {{ISO~26262}},
	year = {2018},
	langid= {english},
}

@book{rausand_risk_2020,
	address = {Hoboken, NJ, USA},
	edition = {2.~Aufl},
	series = {Statistics in {Practice}},
	title = {Risk {Assessment}~{\textendash} {Theory}, {Methods}, and {Applications}},
	isbn = {978-1-119-37722-1 978-1-119-37728-3},
	shorttitle = {Risk assessment},
	publisher = {Wiley},
	author = {Rausand, Marvin and Haugen, Stein},
	year = {2020},
	doi = {10.1002/9781119377351},
	langid= {english},
}

@book{dick_requirements_2017,
	address = {Cham},
	edition = {4.~Aufl},
	title = {Requirements {Engineering}},
	isbn = {978-3-319-61072-6 978-3-319-61073-3},
	url = {http://link.springer.com/10.1007/978-3-319-61073-3},
	language = {en},
	urldate = {2020-04-23},
	publisher = {\foreignlanguage{ngerman}{Springer}},
	author = {Dick, Jeremy and Hull, Elizabeth and Jackson, Ken},
	year = {2017},
	doi = {10.1007/978-3-319-61073-3},
	langid= {english},
}

@book{leveson_engineering_2012,
	address = {Cambridge, MA, USA},
	title = {Engineering a {Safer} {World} {\textendash} {Systems} {Thinking} {Applied} to {Safety}},
	isbn = {978-0-262-29824-7},
	shorttitle = {Engineering a {Safer} {World}},
	language = {en},
	publisher = {MIT Press},
	author = {Leveson, Nancy G.},
	year = {2012},
	doi = {10.7551/mitpress/8179.001.0001},
	langid= {english},
}

@incollection{winner_human_2024,
	address = {Wiesbaden},
	title = {Human {Factors}: {Level} 3+},
	isbn = {978-3-658-38485-2 978-3-658-38486-9},
	shorttitle = {Human {Factors}},
	url = {https://link.springer.com/10.1007/978-3-658-38486-9_38},
	language = {de},
	urldate = {2025-02-24},
	booktitle = {Handbuch {Assistiertes} und {Automatisiertes} {Fahren}},
	publisher = {Springer},
	author = {Bengler, Klaus and Josten, Johanna and Marberger, Claus},
	editor = {Winner, Hermann and Dietmayer, Klaus and Eckstein, Lutz and Jipp, Meike and Maurer, Markus and Stiller, Christoph},
	year = {2024},
	doi = {10.1007/978-3-658-38486-9_38},
	series = {ATZ/\allowbreak{}MTZ-Fachbuch},
	pages = {1021--1034},
	edition = {4.~Aufl},
}

@incollection{winner_nutzergerechte_2024,
	address = {Wiesbaden},
	title = {Nutzergerechte {Gestaltung} der {Mensch}-{Maschine}-{Interaktion} von {Fahrerassistenzsystemen}},
	isbn = {978-3-658-38485-2 978-3-658-38486-9},
	url = {https://link.springer.com/10.1007/978-3-658-38486-9_26},
	language = {de},
	urldate = {2025-02-24},
	booktitle = {Handbuch {Assistiertes} und {Automatisiertes} {Fahren}},
	publisher = {Springer},
	author = {Bengler, Klaus and Eckstein, Lutz},
	editor = {Winner, Hermann and Dietmayer, Klaus and Eckstein, Lutz and Jipp, Meike and Maurer, Markus and Stiller, Christoph},
	year = {2024},
	doi = {10.1007/978-3-658-38486-9_26},
	series = {ATZ/\allowbreak{}MTZ-Fachbuch},
	pages = {671--687},
	edition = {4.~Aufl},
}

@incollection{winner_allgemeine_2024,
	address = {Wiesbaden},
	title = {Allgemeine rechtliche {Rahmenbedingungen} f{\"u}r assistierte, automatisierte und autonome {Fahrfunktionen} in {Deutschland}},
	isbn = {978-3-658-38485-2 978-3-658-38486-9},
	url = {https://link.springer.com/10.1007/978-3-658-38486-9_3},
	language = {de},
	urldate = {2025-01-28},
	booktitle = {Handbuch {Assistiertes} und {Automatisiertes} {Fahren}},
	publisher = {Springer},
	author = {Gasser, Tom Michael},
	editor = {Winner, Hermann and Dietmayer, Klaus and Eckstein, Lutz and Jipp, Meike and Maurer, Markus and Stiller, Christoph},
	year = {2024},
	doi = {10.1007/978-3-658-38486-9_3},
	series = {ATZ/\allowbreak{}MTZ-Fachbuch},
	pages = {37--53},
	edition = {4.~Aufl},
}

@incollection{winner_rahmenbedingungen_2024,
	address = {Wiesbaden},
	title = {Rahmenbedingungen f{\"u}r {Fahrerassistenz} aus {Typgenehmigung} und {Verbraucherschutz}},
	isbn = {978-3-658-38485-2 978-3-658-38486-9},
	url = {https://link.springer.com/10.1007/978-3-658-38486-9_4},
	language = {de},
	urldate = {2025-02-24},
	booktitle = {Handbuch {Assistiertes} und {Automatisiertes} {Fahren}},
	publisher = {Springer},
	author = {Seiniger, Patrick and Seeck, Andre and Wiggerich, Andr{\'e}},
	editor = {Winner, Hermann and Dietmayer, Klaus and Eckstein, Lutz and Jipp, Meike and Maurer, Markus and Stiller, Christoph},
	year = {2024},
	doi = {10.1007/978-3-658-38486-9_4},
	series = {ATZ/\allowbreak{}MTZ-Fachbuch},
	pages = {55--69},
	edition = {4.~Aufl},
}

@book{ross_safety_2021,
	address = {Cham},
	title = {Safety for {Future} {Transport} and {Mobility}},
	publisher = {\foreignlanguage{ngerman}{Springer}},
	author = {Ross, Hans-Leo},
	year = {2021},
	doi = {10.1007/978-3-030-54883-4},
	langid= {english},
}

@techreport{verein_deutscher_ingenieure_entwicklung_2019,
	type = {{Verein Deutscher Ingenieure} {Richtlinie}},
	title = {Entwicklung technischer {Produkte} und {Systeme}},
	number = {{VDI}~2221:2019-11},
	author = {{VDI~2221}},
	year = {2019},
}

@incollection{hoppner_ag--schnittstelle_2008,
	address = {Berlin, Heidelberg},
	title = {{AG}-/{AN}-{Schnittstelle}~{\textendash} {Schwerpunkt} {Aus\-schrei\-bun\-gen}/\allowbreak{}{Ver\-trags\-we\-sen}},
	booktitle = {Das {V}-{Modell}~{XT}},
	publisher = {Springer},
	author = {H{\"o}ppner, Stephan},
	year = {2008},
	doi = {10.1007/978-3-540-30250-6_3},
	series = {eXamen.press},
	pages = {173--274},
}

@techreport{deutsches_institut_fur_normung_din_2020,
	type = {{Deutsches Institut f{\"u}r Normung} Norm},
	title = {{Ergonomie} der {Mensch}-{System}-{Interaktion}~{\textendash} {Teil}~210: {Menschzentrierte} {Gestaltung} interaktiver {Systeme}},
	number = {{DIN} {EN} {ISO}~9241-210:2020-03},
	author = {{DIN EN ISO~9241-210}},
	year = {2020},
}

@incollection{sottilare_towards_2020,
	address = {Cham},
	title = {Towards {Iteration} by {Design}: {An} {Interaction} {Design} {Concept} for {Safety} {Critical} {Systems}},
	volume = {12214},
	booktitle = {Adaptive {Instructional} {Systems}. {HCII}~2020},
	publisher = {\foreignlanguage{ngerman}{Springer}},
	author = {Witte, Thomas E. F. and Hasbach, Jonas and Schwarz, Jessica and Nitsch, Verena},
	editor = {Sottilare, Robert A. and Schwarz, Jessica},
	year = {2020},
	doi = {10.1007/978-3-030-50788-6_17},
	series = {Lecture {Notes} in {Computer} {Science}},
	pages = {228--241},
	langid= {english},
}

@incollection{winner_architektursichten_2024,
	address = {Wiesbaden},
	title = {Architektursichten f{\"u}r {Fahrzeugautomatisierungssysteme}},
	isbn = {978-3-658-38485-2 978-3-658-38486-9},
	url = {https://link.springer.com/10.1007/978-3-658-38486-9_39},
	language = {de},
	urldate = {2025-03-20},
	booktitle = {Handbuch {Assistiertes} und {Automatisiertes} {Fahren}},
	publisher = {Springer},
	author = {Nolte, Marcus and Maurer, Markus},
	editor = {Winner, Hermann and Dietmayer, Klaus and Eckstein, Lutz and Jipp, Meike and Maurer, Markus and Stiller, Christoph},
	year = {2024},
	doi = {10.1007/978-3-658-38486-9_39},
	series = {ATZ/\allowbreak{}MTZ-Fachbuch},
	pages = {1035--1076},
	edition = {4.~Aufl},
}

@techreport{international_organization_for_standardization_isodis_2021,
	type = {{International Organization for Standardization} \foreignlanguage{ngerman}{Norm-Entwurf}},
	title = {{Road} vehicles~{\textemdash} {Safety} of the intended functionality},
	number = {{ISO}/\allowbreak{}{DIS}~21448:2021},
	author = {{ISO/\allowbreak{}DIS~21448}},
	year = {2021},
	langid= {english},
}

@misc{nolte_supporting_2020,
	title = {Supporting {Safe} {Decision} {Making} {Through} {Holistic} {System}-{Level} {Representations}~\& {Monitoring}~{\textendash} {A} {Summary} and {Taxonomy} of {Self}-{Representation} {Concepts} for {Automated} {Vehicles}},
	organization = {arXiv},
	author = {Nolte, Marcus and Jatzkowski, Inga and Ernst, Susanne and Maurer, Markus},
	year = {2020},
	note = {arXiv:2007.13807v2 [eess.SY]},
	doi = {10.48550/ARXIV.2007.13807},
	langid= {english},
}

@incollection{maurer_wirtschaft_2005,
	address = {Berlin, Heidelberg},
	title = {Wirtschaft und gesellschaftliche {Akzeptanz}: {Fahrerassistenzsysteme} auf dem {Pr{\"u}fstand}},
	booktitle = {Fahrerassistenzsysteme mit maschineller {Wahrnehmung}},
	publisher = {Springer},
	author = {Homann, K.},
	editor = {Maurer, Markus and Stiller, Christoph},
	year = {2005},
	doi = {10.1007/3-540-27137-6_11},
	pages = {239--244},
}

@book{wymore_model-based_1993,
	address = {Boca Raton, FL, USA},
	edition = {\foreignlanguage{ngerman}{1.~Aufl}},
	series = {Systems {Engineering} {Series}},
	title = {Model-{Based} {Systems} {Engineering}},
	publisher = {CRC Press},
	author = {Wymore, A. Wayne},
	year = {1993},
	doi = {10.1201/9780203746936},
	langid= {english},
}

@article{forsberg_relationship_1991,
	title = {The {Relationship} of {System} {Engineering} to the {Project} {Cycle}},
	volume = {1},
	doi = {10.1002/j.2334-5837.1991.tb01484.x},
	number = {1},
	journal = {INCOSE International Symposium},
	author = {Forsberg, Kevin and Mooz, Harold},
	year = {1991},
	pages = {57--65},
	langid= {english},
}

@incollection{rios_evolution_2017,
	address = {Cham},
	title = {The {Evolution} of the {V}-{Model}: {From} {VDI}~2206 to a {System} {Engineering} {Based} {Approach} for {Developing} {Cybertronic} {Systems}},
	volume = {517},
	booktitle = {Product {Lifecycle} {Management} and the {Industry} of the {Future}},
	publisher = {\foreignlanguage{ngerman}{Springer}},
	author = {Eigner, Martin and Dickopf, Thomas and Apostolov, Hristo},
	editor = {R{\'i}os, Jos{\'e} and Bernard, Alain and Bouras, Abdelaziz and Foufou, Sebti},
	year = {2017},
	doi = {10.1007/978-3-319-72905-3_34},
	series = {IFIP Advances in Information and Communication Technology},
	pages = {382--393},
	langid= {english},
}

@inproceedings{sifakis_system_2018,
	series = {Electronic {Proceedings} in {Theoretical} {Computer} {Science}},
	title = {System {Design} in the {Era} of {IoT}{\textemdash}{Meeting} the {Autonomy} {Challenge}},
	volume = {272},
	doi = {10.4204/EPTCS.272.1},
	booktitle = {Methods and {Tools} for {Rigorous} {System} {Design} ({MeTRiD} 2018)},
	publisher = {Open Publishing Association},
	author = {Sifakis, Joseph},
	editor = {Bliudze, Simon and Bensalem, Saddek},
	year = {2018},
	pages = {1--22},
	langid= {english},
}

@book{schauffele_automotive_2016,
	address = {Wiesbaden},
	series = {ATZ/\allowbreak{}MTZ-Fachbuch},
	title = {\foreignlanguage{english}{Automotive {Software} {Engineering}}~{\textendash} {Grundlagen}, {Prozesse}, {Methoden} und {Werkzeuge} effizient einsetzen},
	publisher = {Springer},
	author = {Sch{\"a}uffele, J{\"o}rg and Zurawka, Thomas},
	year = {2016},
	doi = {10.1007/978-3-658-11815-0},
	edition = {6.~Aufl},
}

@incollection{eigner_uberblick_2014,
	address = {Berlin, Heidelberg},
	title = {{\"U}berblick {Disziplin}-spezifische und -{\"u}bergreifende {Vorgehensmodelle}},
	booktitle = {Modellbasierte {Virtuelle} {Produktentwicklung}},
	publisher = {Springer},
	author = {Eigner, Martin},
	editor = {Eigner, Martin and Roubanov, Daniil and Zafirov, Radoslav},
	year = {2014},
	doi = {10.1007/978-3-662-43816-9_2},
	pages = {15--52},
}

@inproceedings{wilmsen_method_2020,
	title = {Method for the identification of requirements for designing reference processes},
	doi = {10.1017/dsd.2020.301},
	booktitle = {Proceedings of the 16th {International} {Design} {Conference} ({DESIGN 2020})},
	publisher = {Cambridge University Press},
	author = {Wilmsen, M. and Gericke, K. and J{\"a}ckle, M. and Albers, A.},
	year = {2020},
	pages = {1175--1184},
	langid= {english},
}

@inproceedings{pfeffer_automated_2019,
	address = {Orlando, FL, USA},
	title = {Automated {Driving}~- {Challenges} for the {Automotive} {Industry} in {Product} {Development} with {Focus} on {Process} {Models} and {Organizational} {Structure}},
	isbn = {978-1-5386-8396-5},
	url = {https://ieeexplore.ieee.org/document/8836779/},
	doi = {10.1109/SYSCON.2019.8836779},
	urldate = {2021-05-05},
	booktitle = {2019 {IEEE} {International} {Systems} {Conference}},
	publisher = {IEEE},
	author = {Pfeffer, Raphael and Basedow, Gustav N. and Thiesen, Nina R. and Spadinger, Markus and Albers, Albert and Sax, Eric},
	year = {2019},
	pages = {438--443},
	langid= {english},
}

@article{sexton_effective_2014,
	title = {Effective {Functional} {Safety} {Concept} {Generation} in the {Context} of {ISO} 26262},
	volume = {7},
	issn = {1946-4622},
	url = {http://papers.sae.org/2014-01-0207/},
	doi = {10.4271/2014-01-0207},
	language = {en},
	number = {1},
	urldate = {2017-04-13},
	journal = {SAE International Journal of Passenger Cars - Electronic and Electrical Systems},
	author = {Sexton, Darren and Priore, Antonio and Botham, John},
	year = {2014},
	pages = {95--102},
	langid= {english},
}

@phdthesis{reschka_fertigkeiten-_2017,
	address = {Braunschweig},
	title = {Fertigkeiten- und {F{\"a}higkeitengraphen} als {Grundlage} des sicheren {Betriebs} von automatisierten {Fahrzeugen} im {\"o}ffentlichen {Stra{\ss}enverkehr} in st{\"a}dtischer {Umgebung}},
	language = {de},
	school = {Techn. Univ.},
	author = {Reschka, Andreas},
	collaborator = {{Universit{\"a}tsbibliothek Braunschweig} and Maurer, Markus},
	year = {2017},
	doi = {10.24355/DBBS.084-201707280929},
}

@unpublished{maurer_unfallschwereminderung_2002,
	address = {Essen},
	title = {Unfallschwereminderung durch {Fahrerassistenzsysteme} mit maschineller {Wahrnehmung}~{\textendash} {Potentiale} und {Risiken}},
	author = {Maurer, Markus and W{\"o}rsd{\"o}rfer, Karl-Friedrich},
	year = {2002},
	month = nov,
	howpublished = {Vortrag},
	type = {Seminar Fahrerassistenzsysteme und aktive Sicherheit. Haus der Technik},
}

@incollection{winner_entwicklungsprozess_2015,
	address = {Wiesbaden},
	title = {Entwicklungsprozess von {Kollisionsschutzsystemen} f{\"u}r {Frontkollisionen}: {Systeme} zur {Warnung}, zur {Unfallschwereminderung} und zur {Verhinderung}},
	isbn = {978-3-658-05733-6 978-3-658-05734-3},
	shorttitle = {Entwicklungsprozess von {Kollisionsschutzsystemen} f{\"u}r {Frontkollisionen}},
	url = {https://link.springer.com/10.1007/978-3-658-05734-3_48},
	language = {de},
	urldate = {2025-02-03},
	booktitle = {Handbuch {Fahrerassistenzsysteme}},
	publisher = {Springer},
	author = {Reschka, Andreas and Rieken, Jens and Maurer, Markus},
	editor = {Winner, Hermann and Hakuli, Stephan and Lotz, Felix and Singer, Christina},
	year = {2015},
	doi = {10.1007/978-3-658-05734-3_48},
	series = {ATZ/\allowbreak{}MTZ-Fachbuch},
	pages = {913--935},
	edition = {3.,~{\"u}berarbeitete und erg{\"a}nzte Aufl},
}

@inproceedings{graubohm_systematic_2017,
	address = {\foreignlanguage{ngerman}{M{\"u}nchen}},
	title = {Systematic {Design} of {Automated} {Driving} {Functions} {Considering} {Functional} {Safety} {Aspects}},
	booktitle = {8.~{Tagung} {Fahrerassistenz}},
	publisher = {\foreignlanguage{ngerman}{Techn. Univ. M{\"u}nchen~{\textendash} Lehrstuhl f{\"u}r Fahrzeugtechnik}},
	author = {Graubohm, Robert and Stolte, Torben and Bagschik, Gerrit and Reschka, Andreas and Maurer, Markus},
	year = {2017},
	langid= {english},
}

@techreport{international_organization_for_standardization_isotr_2020,
	type = {{International Organization for Standardization} \foreignlanguage{ngerman}{Techn. Ber.}},
	title = {{Road} vehicles~{\textemdash} {Safety} and cybersecurity for automated driving systems~{\textemdash} {Design}, verification and validation},
	number= {{ISO}/{TR}~4804:2020},
	author = {{ISO/\allowbreak{}TR~4804}},
	year = {2020},
	langid= {english},
}

@article{boehm_spiral_1988,
	title = {A {Spiral} {Model} of {Software} {Development} and {Enhancement}},
	volume = {21},
	url = {http://ieeexplore.ieee.org/document/59/},
	doi = {10.1109/2.59},
	language = {en},
	number = {5},
	urldate = {2020-04-16},
	journal = {Computer},
	author = {Boehm, Barry W.},
	year = {1988},
	pages = {61--72},
	langid= {english},
}

@misc{beck_manifesto_2001,
	title = {Manifesto for {Agile} {Software} {Development}},
	url = {https://agilemanifesto.org/},
	urldate = {2021-05-05},
	howpublished  = {Agile Alliance},
	author = {Beck, Kent and Beedle, Mike and {van Bennekum}, Arie and Cockburn, Alistair and Cunningham, Ward and Fowler, Martin and Grenning, James and Highsmith, Jim and Hunt, Andrew and Jeffries, Ron and Kern, Jon and Marick, Brian and Martin, Robert C. and Mellor, Steve and Schwaber, Ken and Sutherland, Jeff and Thomas, Dave},
	year = {2001},
	langid= {english},
}

@article{cooper_process_1983,
	title = {A {Process} {Model} for {Industrial} {New} {Product} {Development}},
	volume = {EM-30},
	url = {https://ieeexplore.ieee.org/document/6448637/},
	doi = {10.1109/TEM.1983.6448637},
	number = {1},
	urldate = {2025-03-05},
	journal = {IEEE Transactions on Engineering Management},
	author = {Cooper, Robert G.},
	year = {1983},
	pages = {2--11},
	langid= {english},
}

@incollection{doorn_value_2013,
	address = {\foreignlanguage{ngerman}{Dordrecht, Niederlande}},
	title = {Value {Sensitive} {Design} and {Information} {Systems}},
	volume = {16},
	isbn = {978-94-007-7843-6 978-94-007-7844-3},
	url = {http://link.springer.com/10.1007/978-94-007-7844-3_4},
	urldate = {2018-07-04},
	booktitle = {Early {Engagement} and {New} {Technologies}: {Opening} {Up} the {Laboratory}},
	publisher = {\foreignlanguage{ngerman}{Springer}},
	author = {Friedman, Batya and Kahn, Peter H. and Borning, Alan and Huldtgren, Alina},
	editor = {Doorn, Neelke and Schuurbiers, Daan and {van de Poel}, Ibo and Gorman, Michael E.},
	year = {2013},
	doi = {10.1007/978-94-007-7844-3_4},
	series = {Philosophy of Engineering and Technology},
	pages = {55--95},
	langid= {english},
}

@inproceedings{graubohm_value_2020,
	title = {Value {Sensitive} {Design} in the {Development} of {Driverless} {Vehicles}: {A} {Case} {Study} on an {Autonomous} {Family} {Vehicle}},
	doi = {10.1017/dsd.2020.140},
	booktitle = {Proceedings of the 16th {International} {Design} {Conference} ({DESIGN 2020})},
	publisher = {Cambridge University Press},
	author = {Graubohm, Robert and Schr{\"a}der, Tobias and Maurer, Markus},
	year = {2020},
	pages = {907--916},
	langid= {english},
}

@inproceedings{ohl_autonomes_2012,
	address = {Braunschweig},
	title = {Autonomes {Fahren} im {Stra{\ss}enbetriebsdienst} auf {Autobahnen}},
	booktitle = {{AAET} 2012~{\textendash} {Automatisierungssysteme}, {Assistenzsysteme} und eingebettete {Systeme} für {Transportmittel}},
	publisher = {ITS mobility~e.V.},
	author = {Ohl, Sebastian and Maurer, Markus and H{\"a}usler, Katharina and Holldorb, Christian},
	year = {2012},
	pages = {252--272},
}

@inproceedings{stolte_towards_2015,
	address = {\foreignlanguage{ngerman}{Las Palmas de Gran Canaria, Spanien}},
	title = {Towards {Automated} {Driving}: {Unmanned} {Protective} {Vehicle} for {Highway} {Hard} {Shoulder} {Road} {Works}},
	doi = {10.1109/ITSC.2015.115},
	booktitle = {2015 {IEEE} 18th {International} {Conference} on {Intelligent} {Transportation} {Systems}},
	publisher = {IEEE},
	author = {Stolte, Torben and Reschka, Andreas and Bagschik, Gerrit and Maurer, Markus},
	year = {2015},
	pages = {672--677},
	langid= {english},
}

@inproceedings{graubohm_functional_2019,
	address = {\foreignlanguage{ngerman}{Delft, Niederlande}},
	title = {Functional {Safety} {Concept} {Generation} within the {Process} of {Preliminary} {Design} of {Automated} {Driving} {Functions} at the {Example} of an {Unmanned} {Protective} {Vehicle}},
	doi = {10.1017/dsi.2019.293},
	booktitle = {Proceedings of the 22nd {International} {Conference} on {Engineering} {Design} ({ICED19})},
	publisher = {Cambridge University Press},
	author = {Graubohm, Robert and Stolte, Torben and Bagschik, Gerrit and Steimle, Markus and Maurer, Markus},
	year = {2019},
	pages = {2863--2872},
	langid= {english},
}

@inproceedings{stolte_hazard_2017,
	address = {Redondo Beach, CA, USA},
	title = {Hazard {Analysis} and {Risk} {Assessment} for an {Automated} {Unmanned} {Protective} {Vehicle}},
	isbn = {978-1-5090-4804-5},
	url = {http://ieeexplore.ieee.org/document/7995974/},
	doi = {10.1109/IVS.2017.7995974},
	urldate = {2018-01-12},
	booktitle = {2017 {IEEE} {Intelligent} {Vehicles} {Symposium}},
	publisher = {IEEE},
	author = {Stolte, Torben and Bagschik, Gerrit and Reschka, Andreas and Maurer, Markus},
	year = {2017},
	pages = {1848--1855},
	langid= {english},
}

@misc{jentzsch_man_2018,
	title = {{MAN} {aFAS} erh{\"a}lt den ersten {\quotesinglbase}{Truck} {Innovation} {Award}{\textquoteleft} auf der {IAA} {Nutzfahrzeuge} 2018},
	url = {https://press.mantruckandbus.com/corporate/de/man-afas-erhalt-den-ersten-truck-innovation-award-auf-der-iaa-nutzfahrzeuge-2018/},
	urldate = {2021-05-05},
	howpublished = {MAN \foreignlanguage{english}{Newsroom Corporate}},
	author = {Jentzsch, Gregor},
	month = sep,
	year = {2018},
}

@unpublished{bagschik_funktionale_2018,
	address = {Frankfurt am Main},
	title = {Funktionale {Sicherheit} f{\"u}r das automatisch fahrerlos fahrende {Absicherungsfahrzeug}},
	author = {Bagschik, Gerrit and Stolte, Torben and Steimle, Markus and Maurer, Markus},
	month = jun,
	year = {2018},
	howpublished = {Vortrag},
	type = {aFAS Schlusspräsentation. \foreignlanguage{english}{House of Logistics \& Mobility}},
}
\end{document}